\documentclass[11pt]{article}
\usepackage{graphicx,tabularx,epsfig}
\usepackage{color}
\usepackage{enumitem}
\usepackage{caption}
\usepackage{subcaption}
\usepackage{amsmath}
\usepackage{amssymb}
\usepackage{amsthm}
\usepackage{physics}
\usepackage{dsfont, mathtools}
\usepackage{psvectorian}
\usepackage{blkarray}
\usepackage{bm}
\usepackage{url}
\usepackage{setspace}
\usepackage{xcolor}
\colorlet{linkequation}{blue}

\usepackage[colorlinks = true,
            linkcolor = blue,
            urlcolor  = blue,
            citecolor = blue,
            anchorcolor = blue]{hyperref}

\usepackage{floatrow}

\numberwithin{equation}{section}

\newlength{\abstractwidth}
\renewcommand{\thefootnote}{\fnsymbol{footnote}}
\renewcommand{\thanks}[1]{\footnote{#1}} % Use this for footnotes
\newcommand{\starttext}{
\setcounter{footnote}{0}
\renewcommand{\thefootnote}{\arabic{footnote}}}

\makeatletter
\g@addto@macro\normalsize{%
  \setlength\abovedisplayskip{15pt}
  \setlength\belowdisplayskip{15pt}
  \setlength\abovedisplayshortskip{15pt}
  \setlength\belowdisplayshortskip{15pt}
}
\makeatother

\usepackage{color}

\renewcommand{\title}[1]{\vbox{\center\LARGE{#1}}\vspace{5mm}}
\renewcommand{\author}[1]{\vbox{\center#1}\vspace{5mm}}

\begin{document}

\singlespacing

\begin{center}

{\Large \bf {Bouncing inside the horizon and scrambling delays}}

\bigskip \noindent
\bigskip
\bigskip

 { Gary T. Horowitz$^a$, Henry Leung$^a$, Leonel Queimada$^a$, and Ying Zhao$^b$}

\bigskip
\bigskip

    {\it $^a$  Department of Physics, University of California, Santa Barbara, CA 93106}
    \vskip1em
   {\it $^b$ Kavli Institute for Theoretical Physics,
Santa Barbara, CA 93106}

\bigskip
\bigskip
    
 %   {\tt email}

\bigskip
\bigskip
\bigskip

\end{center}

\begin{abstract}

We study charged perturbations of the thermofield double state dual to a charged AdS black hole. We model the perturbation by a massless charged shell in the bulk. Unlike the neutral case, all such shells bounce at a definite radius, which can be behind the horizon. We show that the standard ``shock wave" calculation of a scrambling time indicates that adding charge  increases the scrambling time. We then give two arguments using the bounce that suggest that scrambling does not actually take longer when charge is added, but  instead its onset is delayed. We also construct a boundary  four point function  which detects whether the shell  bounces  inside the black hole.

\medskip
\noindent
\end{abstract}

\newpage

\starttext \baselineskip=17.63pt \setcounter{footnote}{0}

{\hypersetup{hidelinks}
\tableofcontents
}

\section{Introduction}
It has been shown that the physics near a black hole horizon is closely related to quantum chaos \cite{Sekino:2008he}. In particular, the process of a particle falling into an asymptotically 
anti-de Sitter (AdS) black hole corresponds to scrambling of the thermofield double state in the dual field theory \cite{Shenker:2013pqa, Shenker:2013yza}. 
However the dual interpretation of the trajectory of a particle inside the black hole  is much less understood. 

\begin{figure}[H] 
 \begin{center}                     
      \includegraphics[width=1.4in]{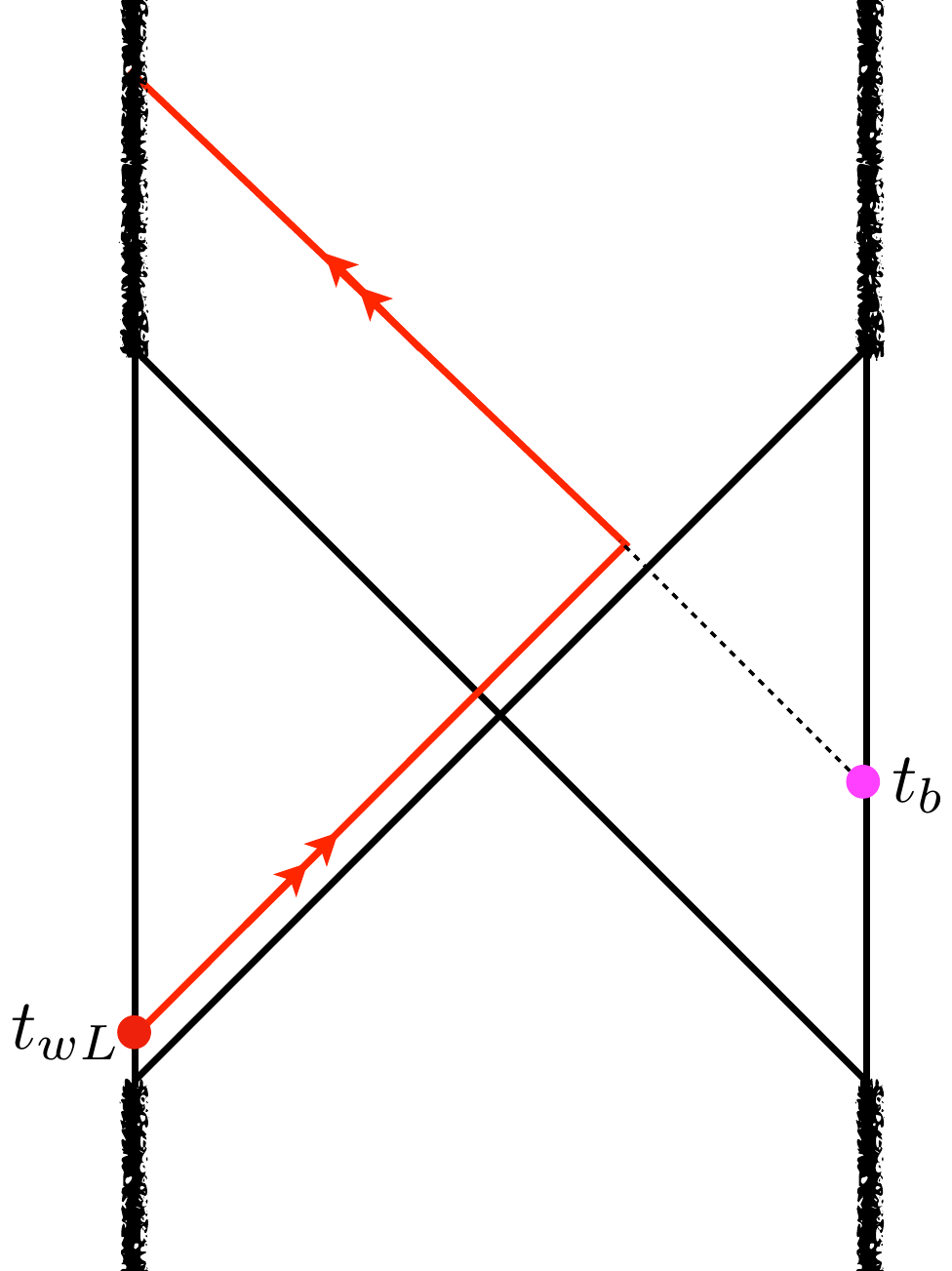}
      \caption{A charged shell sent in from the left at time $t_{wL}$ can bounce in the interior close to the event horizon. An observer on the right can meet the shell only if they fall in before time $t_b$. }
  \label{trajectory_bouncing_intro}
  \end{center}
\end{figure}

To shed light on this issue,  we study a charged perturbation of  the  state dual to a  (four-dimensional) charged black hole. We model the bulk perturbation by a  spherical, massless charged shell. As we will review, all such shells bounce at a certain radius, so a shell that is sent in from the left asymptotic region ends on the left side of the Penrose diagram. By choosing the energy and charge of the shell appropriately, we can arrange for the shell to bounce inside the black hole, but close to the event horizon (see Fig. \ref{trajectory_bouncing_intro}).

We first calculate a scrambling time for the charged perturbation by generalizing the original calculation of scrambling for a neutral perturbation of a neutral black hole
\cite{Shenker:2013pqa, Shenker:2013yza}. This was extended to a neutral perturbation of a charged black hole in \cite{Leichenauer:2014nxa}, and we further extend it to a charged perturbation of a charged black hole.\footnote{This was previously studied in \cite{Reynolds:2016pmi} in three dimensions, even though the presence of the bounce was not considered.} We find that the scrambling time appears to increase with the addition of charge.
 This calculation does not probe the shell inside the black hole, and hence seems independent of the bounce. However, we then point out a surprising coincidence. Let $t_{wL}$ be the time the charged shell is sent in from the left. Since the shell bounces inside the horizon, there is a maximum time $t_b$ at which an observer on the right can fall in and still meet the shell (see Fig. \ref{trajectory_bouncing_intro}). We show that, away from extremality, the difference in scrambling times between a charged and neutral shell with the same energy is just\footnote{We define $t$ so that it increases to the future on both asymptotic boundaries. So in Fig. \ref{trajectory_bouncing_intro}, $t_{wL} < 0$. }  $-t_{wL}- t_b$  (up to an order one multiple of the thermal time).

We then give two arguments which depend on the physics inside the horizon that indicate that, from one point of view, the actual scrambling time does not increase with charge. Instead, the effect of the charge is to delay the start of scrambling. The first is motivated by considering quantum circuits. It has been argued that the geometry on particular spacelike surfaces in the bulk reflects the minimal tensor network required to prepare the state \cite{Hartman:2013qma}. Since these surfaces extend inside the black hole, one can model evolution inside it in terms of a unitary quantum circuit. A description of black hole scrambling has been developed using these quantum circuits \cite{Haehl:2021emt, Zhao:2020gxq}.

Let us now consider two perturbations (one sent in from the left and one from the right) that meet inside the black hole. 
It was proposed in \cite{Haehl:2022uop} that one can compare the scrambling times of these two perturbations by computing the four-volume of the post-collision region, and seeing how it varies as one changes the time at which one perturbation is introduced. We construct the solution with two null shells, one charged and one neutral, and compute the volume of the post-collision region. The result does not behave as expected if the two scrambling times are different. Instead, it behaves exactly as predicted if the scrambling times are the same, but the onset of scrambling for the charged perturbation is delayed.  

Our second argument for the delay comes from a calculation in a two-dimensional black hole. A massless charged particle bounces inside the black hole just like in higher dimensions. We compute an out-of-time-order  four-point correlator (OTOC) with two charged and two neutral operators. With one choice of times for the operator insertions we obtain an ``exterior OTOC" which depends only on the perturbation outside the black hole. This is similar to the OTOC that was studied in the seminal work \cite{Shenker:2013pqa}. With a different choice of times, we obtain an ``interior OTOC" which is sensitive to  the perturbation inside the horizon. We find that the interior OTOC only starts to decay after a certain delay. 

When the operators are all neutral, a certain derivative of this OTOC was shown to reproduce the momentum of a bulk particle (in a particular frame) \cite{Lin:2019qwu}. We show that with two charged operators, the same derivative of the interior OTOC vanishes at the time when the charged particle bounces. Independent of scrambling, it is interesting  that the interior OTOC provides a boundary observable that detects the bounce of a charged particle inside the black hole.

\section{Bouncing trajectories}
\label{sec:bounce_inside_bulk}

In this section, we first review the motion of charged null shells falling into a charged black hole.
 We will see that they always bounce, and with appropriately chosen charge, the bounce will occur inside the black hole, but close to the event horizon.  We then compute a scrambling time for the dual charged perturbation of the black hole state, by generalizing the work of \cite{Shenker:2013pqa, Leichenauer:2014nxa, Reynolds:2016pmi}. We will see that the scrambling time computed this way 
  is longer for a charged shell than a neutral shell with the same energy.

\subsection{Geometry and trajectories of charged shells}

We start with the four-dimensional asymptotically AdS Reissner-Nordstrom spacetime
\begin{equation}\label{RN}
ds^2=-f(r)dt^2+f(r)^{-1}dr^2+r^2 d\Omega^2\,,
\end{equation}
with 
\begin{equation}
\label{fr}
f(r)=\frac{r^2}{l^2}+1-\frac{2M}{r}+\frac{Q^2}{r^2}\,,
\end{equation}
where $M$ is the mass of the black hole and $Q$ is its charge. The horizons of the black hole are given by $f(r_{\pm})=0$. It will also be convenient to use Kruskal coordinates defined through 
\begin{equation}\label{Kruskal_1}
UV=-e^{4\pi r^{*}/ \beta}\,,
\end{equation}
\begin{equation}\label{Kruskal_2}
\frac{U}{V}=-e^{-4\pi t/ \beta}\,,
\end{equation}
with the tortoise coordinate $r^{*}$ defined as 
\begin{equation}\label{tortoise}
r^*(r)=-\int_r ^\infty \frac{dr'}{f(r')}\,,
\end{equation}
so that $r^{*}=0$ at the boundary.

We now add a massless charged shell with energy $E$ and charge $q$.
 The resulting solution is given by a black hole geometry with mass $M$ and charge $Q$ inside the shell, glued to a black hole with mass $M+E$ and charge $Q+q$ outside the shell.  The surface between the two regions is the trajectory of the shell. Therefore, the geometry inside the shell and outside will have a metric given by \eqref{RN} with different choices of $f(r)$, which we will denote by $f_i(r)$ and $f_o(r)$, respectively.

\begin{figure}[H] 
 \begin{center}                      
      \includegraphics[width=4in]{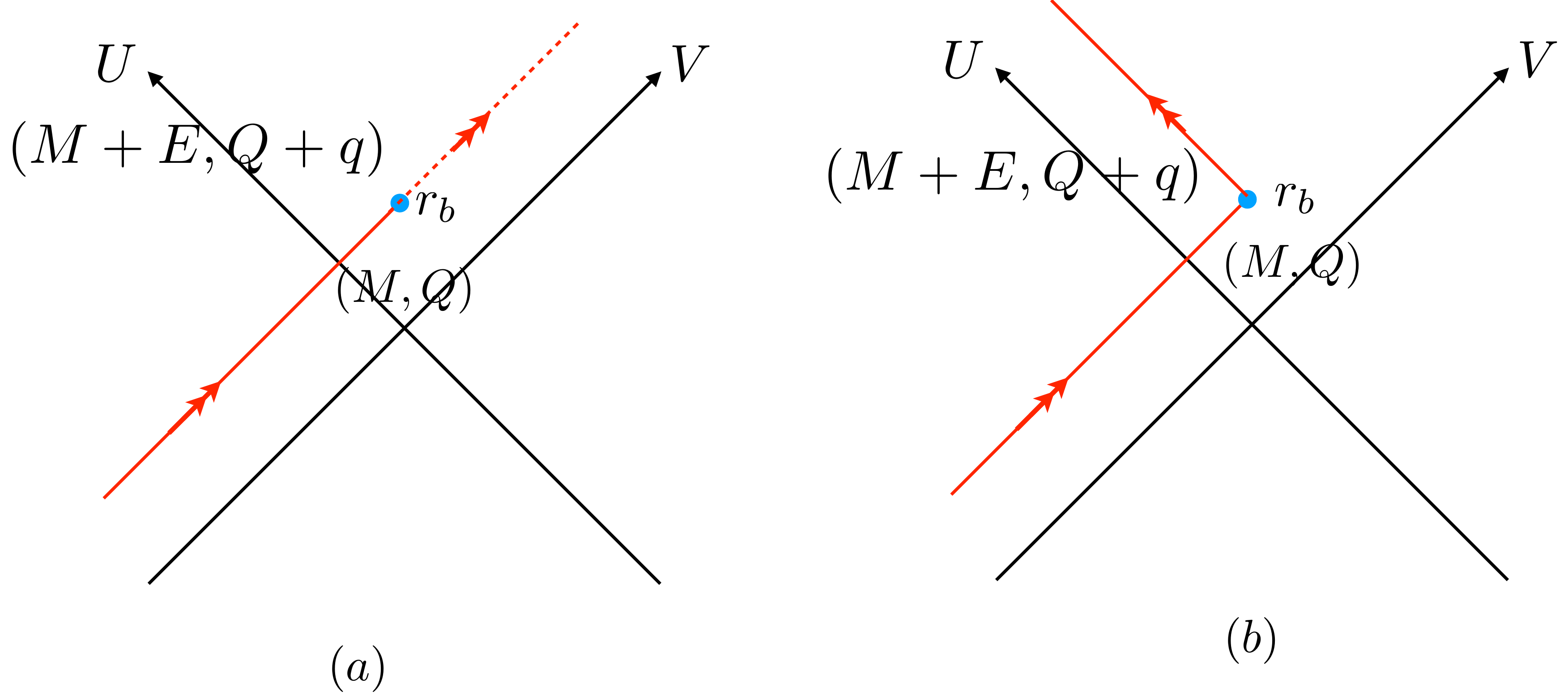}
      \caption{(a) Null energy condition is violated beyond $r_b$. (b) Correct trajectory of a charged shell.}
  \label{coord}
  \end{center}
\end{figure}

One may consider the gluing using the null junction formalism \cite{PhysRevD.43.1129}, where one specifies the null surface and calculates the stress tensor of the shell. In particular, the conditions require that the degenerate induced metric on the null surface must be continuous and the surface stress tensor is related to the discontinuity in the derivatives of the metric. The naive choice for a surface representing an in-going shell sent from the left boundary is to use a constant $U$ surface, i.e. the same surface that represents the trajectory of a neutral shell in \cite{Shenker:2013pqa, Leichenauer:2014nxa}. One finds the surface stress tensor for this shell to be of the form
\begin{equation}
    S^{\alpha \beta}=\rho(r) k^{\alpha} k^{\beta}\,,
\end{equation}
with 
\begin{equation}
    \rho(r)=\frac{f_i(r)-f_o(r)}{8\pi r},
\end{equation}
and $k^\alpha$ is the tangent vector along the shockwave.
This stress tensor vanishes at the radius
\begin{equation}\label{rt}
    r_b=\frac{q(2Q+q)}{2E},
\end{equation}
where $f_i(r_b)= f_o(r_b)$.
If one extends the trajectory to smaller radii, one finds $\rho(r) < 0$ so the stress energy tensor  would violate the null energy condition ($T_{\mu\nu} \ell^\mu \ell^\nu \ge 0$ for all null $\ell$). This problem was addressed in \cite{Dray_1990,Ori:1991} where it was shown that the correct evolution of the matter past $r_b$ is that the in-going particles  reflect and become out-going particles. Thus, at $r_b$ the charged shell will cease to follow the $U$ constant trajectory and start following a constant $V$ surface (see Fig. \ref{coord}).

A good way to understand this bounce is to think of the massless shell as the limit of a massive shell. As we show in Appendix \ref{app:massive}, a massive charged shell always bounces since it is attracted to the oppositely charged singularity. As the mass of the shell goes to zero, the bounce becomes sharper and sharper, leading to the reflection seen in the massless case.

Another way to derive the location of the bounce is to use the DTR conditions \cite{Dray:1985yt, 10.1143/PTP.73.1401, PhysRevD.41.1796}  for matching across two colliding null shells. In the limit when the second shell disappears, i.e., its energy goes to zero, the condition reduces to $f_i(r_b) = f_o(r_b)$.

 In the limit where the shell is a small perturbation with $q\ll Q$, the turning point simplifies to $r_b\approx q Q/E$. In this case,  the turning point only exists when the charge of the black hole $Q$ and the charge of the shell $q$ are of the same sign, as one might expect. At energy $E=q Q/r_+$, the shell bounces exactly at the event horizon.   In fact, this is the energy that leads to a vanishing variation of entropy from the first law
\begin{equation}
    TdS=dM-\Phi dQ=E-\frac{q Q}{r_+}.
\end{equation}
This also shows that shells that bounce inside the black hole always lead to $\Delta S>0$ in agreement with the second law, and shells bounce outside if and only if the parameters of the shell are such that they would lead to $\Delta S<0$ if the shell were to enter the black hole.\footnote{ This sharp statement is only true for massless shells. For massive shells, it is possible for shells to have parameters that would lead to $\Delta S>0$ but still bounce outside the horizon. } 

If the charge on the shell is large enough so that the bounce occurs outside the horizon, then a field theory description of the black hole interacting with a massless, charged scalar field typically exhibits instabilities. This is because $\omega < qQ/r_+$ is the condition for charged superradiance, so the black hole develops scalar hair
 \cite{Dias:2016pma}. In this paper we will focus on the case of bouncing inside the horizon as shown in Figure \ref{coord}(b). The scrambling effects we are interested in will be large when
  the bounce occurs close to the event horizon, i.e., $r_b > r_-$ and $r_+ - r_b \ll r_+$.

\subsection{Butterfly effect with charge}
\label{sec:scrambling time}
In the previous section, we obtained the geometry of general massless charged shells. We will now restrict to the case of a small perturbation, where $E\ll M$ and $q\ll Q$. If we simply take the limit $E,q\to 0$, the backreaction on the geometry by the shell vanishes and the equation of motion of the shell  reduces to the equation for a collection of test particles. However, if one simultaneously takes the limit that the time, $t_{wL}$, at which the perturbation is made is asymptotically early, then there will still be a nontrivial backreaction effect on the geometry. This implies an effect on the thermofield double state at $t=0$, which can be detected by looking at the change in mutual information or an appropriate OTOC which are related to spacelike probes that anchor to each boundary at $t=0$. 
 This leads to the butterfly effect originally obtained in \cite{Shenker:2013pqa, Leichenauer:2014nxa} and we will consider how the presence of charge modifies those results.

In Appendix \ref{app:scrambling}, we give the 
(by now standard) calculation showing that in the limit $\beta_i/\delta \beta \gg -t_{wL}/\beta_i \gg 1$ the main effect of the backreaction of the shell is a shift of the Kruskal $V$ coordinate:
\begin{equation}
V_o \approx V_i + \alpha
\end{equation}
with 
\begin{equation}\label{eq:alpha}
\alpha=e^{C} e^{-2\pi t_{wL}/\beta}\ \frac{\delta r_{+}}{r_{+}-r_-}\,,
\end{equation}
where $C$ is an order one constant. As mentioned in the previous section, a shell that enters the black hole increases its entropy, hence $\delta r_+>0$ and the shift $\alpha$ is positive. 

Notice that no matter how small $\delta r_+/r_+$ is,  we can make $\alpha\sim 1$ by taking $t_{wL}$ far in the past, so spacelike probes at $t=0$ that cross the shell will experience significant changes and signify a large change in the state at $t=0$ in the presence of a small perturbation. This is the hallmark of chaos. 

We first examine the bounce to see whether it can affect spacelike probes that intersect the shell. The analysis in Appendix \ref{app:scrambling} shows that for a bounce near the horizon, we have \eqref{eq:near_horizon_UV}:
\begin{equation}
V_i ^b = -\left( \frac{r_b-r_{+}}{\delta r_{+}}\right) \alpha \,.
\end{equation}
Finding $\delta r_{+}$ through the first law of thermodynamics:
\begin{equation}
\delta r_{+}=\frac{\beta}{2\pi r_{+}}\left(E-\frac{qQ}{r_{+}}\right)\,
\end{equation}
and using \eqref{rt} for the bounce (in the limit $q\ll  Q$), we find 
\begin{equation}\label{eq:V_b}
V_i ^b = \frac{2\pi r_{+}^2 \alpha}{E\beta}\,.
\end{equation}
When backreaction is relevant, i.e. $\alpha\sim 1$, we always have $|V_i^b|\gg 1$ since $r_+/E\gg 1$ for weak perturbations. The Wheeler-DeWitt patch at $t=0$ is bounded by the null hypersufaces $V_o=- 1$, $U_o= 1$ near the left boundary and $V_i=1$, $U_i=-1$ near the right boundary. Following the hypersurfaces $V_o=- 1$, $U_o= 1$ across the shell and writing them in terms of $V_i,U_i$, one obtains a shift that is at most $O(1)$.  Therefore, shells that have strong backreaction never bounce inside this region and, consequently, spacelike probes at $t=0$ will not reach the turning point of such shells.

In the limit we are working in, the difference between the effects of charged and uncharged shells lies only in $\delta r_+$ in the expression of $\alpha$. In the case of uncharged shells $\delta r_+\propto E$, whereas $\delta r_+\propto E-qQ/r_+$ for charged shells. 
Rather than focus on the effect of the perturbation on the state at $t=0$, we can use the boost symmetry to rephrase the result in terms of the state at any time $t_f$ on the left (and $-t_f$ on the right). The only change is that $-t_{wL}$ in the exponent in (\ref{eq:alpha}) is replaced by $t_f -t_{wL} $.
Now consider the following two processes. At time $t_{wL}$, we can either throw in a shell with energy $E$ and charge $q$ ($qQ>0$), or we can throw in a neutral shell with energy $E$. Let   $t_f$ be the time at which the shell will induce significant backreaction ($\alpha \sim 1$) on the geometry. Then (\ref{eq:alpha}) implies
\begin{align}
\label{diff}
    t_f^{(E, q)}-t_f^{(E,0)} = \frac{\beta}{2\pi}\log\frac{E}{E-Qq/r_+}
\end{align}
i.e., compared with a neutral shell carrying the same energy, the charged shell takes a longer time to induce the same amount of backreaction to the geometry. This suggests that the scrambling time of charged perturbations is longer. Even though the above calculation is independent of the bounce, notice that the difference \eqref{diff}  becomes large when the shell bounces close to the horizon since $E-Qq/r_+$ approaches zero. 

It should be noted that this difference in scrambling times is the same as one would have with two neutral shells  if one had energy $E_1 = E -\mu q $ for some $q$ (where $\mu = Q/r_+$), and the other had energy $E_2 = E$. So from the standpoint of this section, the effect of adding charge is indistinguishable from decreasing the energy. In particular,  the bounce is clearly playing no role. In Sec 3 and Sec. 4 we will discuss effects for which the bounce does play a role.

\subsection{Coincidence}

\begin{figure}[H] 
 \begin{center}                      
      \includegraphics[width=1.5in]{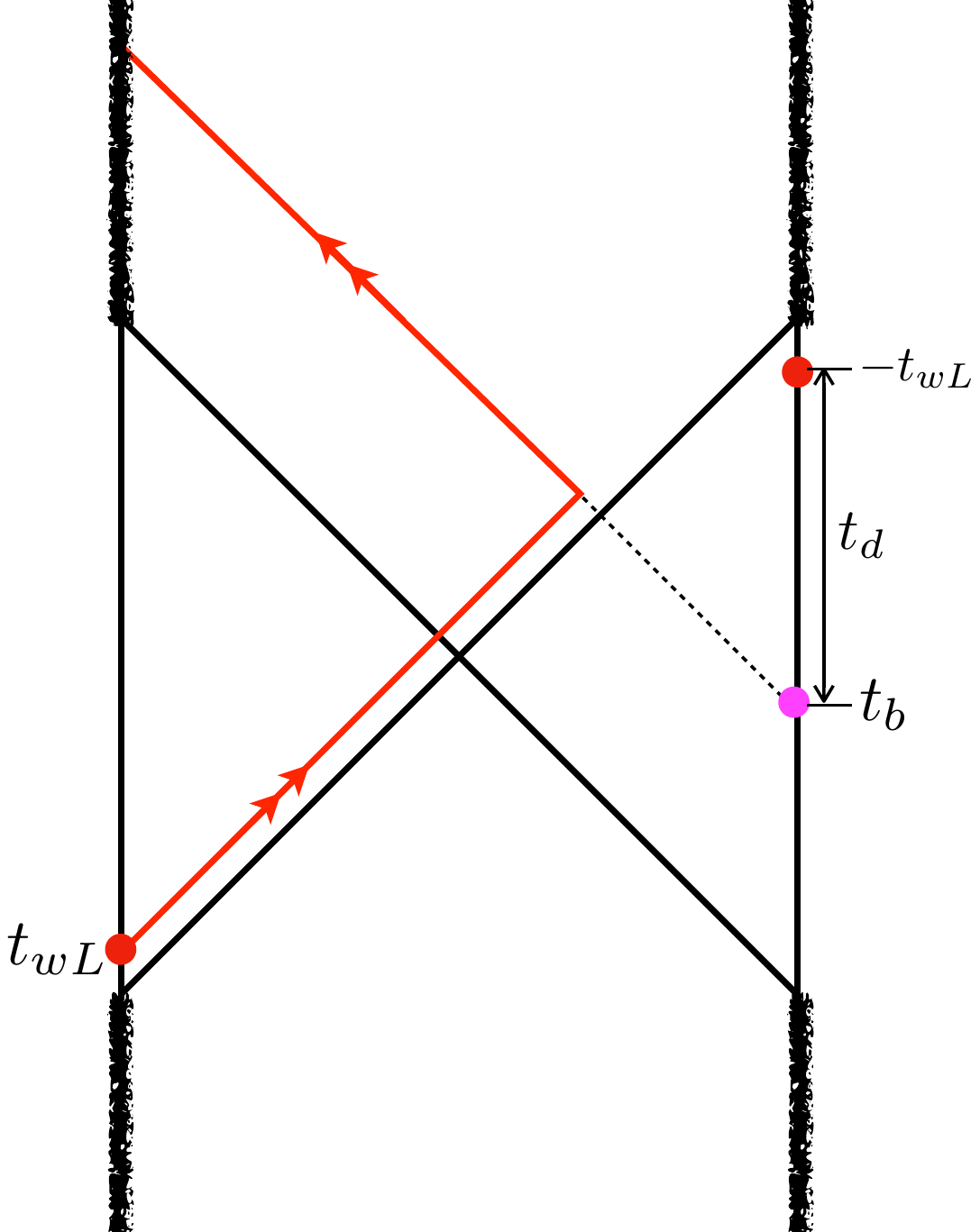}
      \caption{An observer on the right can meet the charged shell only if they fall in before time $t_b$.}
  \label{trajectory_bouncing}
  \end{center}
\end{figure}

In the regime where the above time difference is much larger than the thermal time $\beta$, there is an interesting coincidence. Consider a shell sent from the left boundary at time $t_{wL}$. Since we have chosen $t$ to increase to the future on both boundaries, the corresponding time on the right under analytic continuation is $-t_{wL}$.  A natural way to detect the bouncing of the shell is to note that there is a latest time on the right, $t_b$, for an observer to fall in and still meet the shell. If the observer falls in any later, the shell bounces and the two never intersect (see Fig. \ref{trajectory_bouncing}).  From \eqref{eq:V_b} one  finds the time difference $t_d\equiv-t_{wL}-t_b$ to be
\begin{equation}\label{t_delay1}
t_d = -t_{wL}-t_b=\frac{\beta}{2\pi}\left[\log{\left(\frac{E}{E-qQ/r_+}\cdot\frac{r_+-r_-}{r_+}\right)}-C \right]\,.
\end{equation}

Note that this time difference is a boost invariant quantity. Moreover, recall that $C$ is an $\mathcal{O}(1)$ constant defined in Appendix  \ref{app:scrambling} that depends only on the geometry.  Therefore, away from extremality, the difference in scrambling times $t_f^{(E, q)}-t_f^{(E,0)}$ agrees with $t_d$ up to  terms of order the thermal time.
As we will argue in the next section, one can interpret $t_d$ as the amount of scrambling delay.

\section{Scrambling delay from interior bounces}
\label{sec:collision}

In this section we use a proposed connection between physics inside a black hole and properties of a quantum circuit to  shed light on the difference between scrambling of neutral and charged perturbations. We will find evidence that rather than charged perturbations taking a longer time to scramble (as indicated in the previous section),  the effect of adding charge is to delay the start of scrambling.

\subsection{Collision in the black hole interior and a quantum circuit description}

We begin by briefly reviewing the quantum circuit description of a black hole interior, and how it is related to the interior trajectory of an infalling object as well as collisions inside the black hole.

\begin{figure}[H] 
 \begin{center}         
 \includegraphics[width=3.6in]{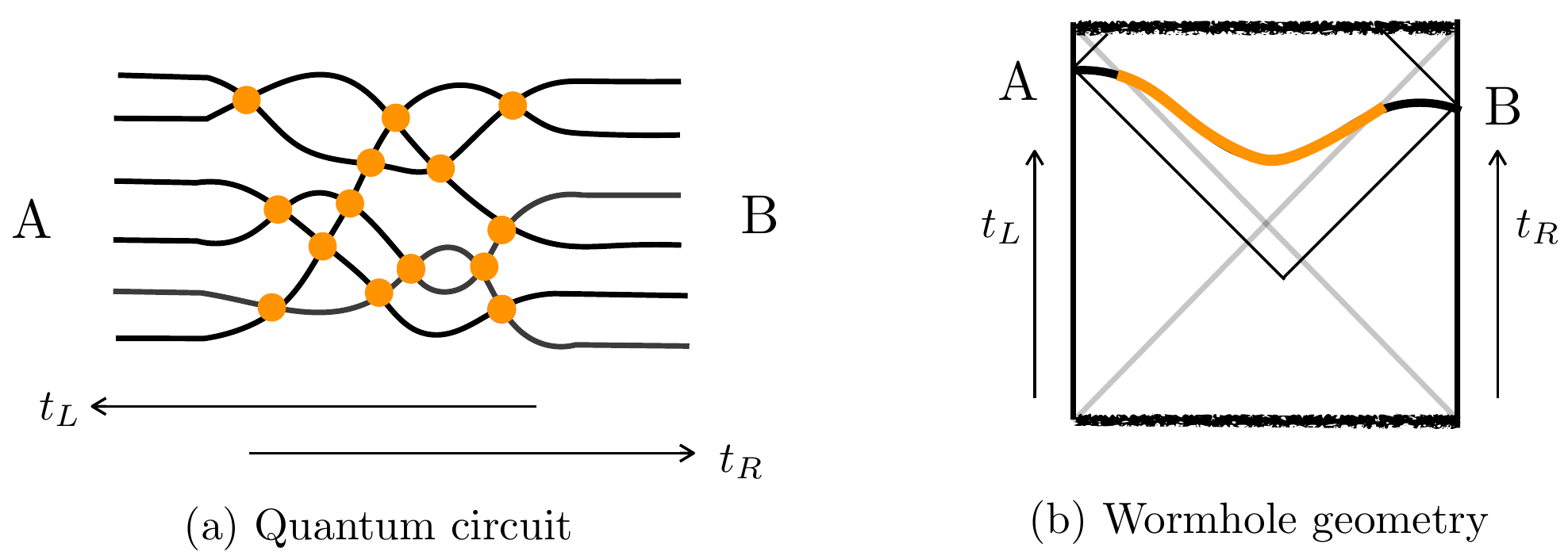}
   \caption{ {\it{Left Panel:}} A unitary quantum circuit with local interactions which becomes longer when $t_L$ or $t_R$ is increased. {\it{Right Panel:}} An extremal surface in a Schwarzschild AdS black hole connecting $t_L$ on the left and $t_R$ on the right. The part of the surface inside the black hole (marked in yellow) increases as $t_L$ or $t_R$ is increased, and is believed to be described by a quantum circuit.} 
     \label{thermofield_double}
  \end{center}
\end{figure}

In the gauge-gravity correspondence, it was argued that the bulk geometry reflects the minimal tensor network preparing the state \cite{Swingle:2009bg}. In particular, there is a unitary quantum circuit that becomes longer as one evolves the state, reflecting the fact that the black hole interior along an extremal surface  gets longer \cite{Hartman:2013qma, Susskind:2014moa} (see Fig.~\ref{thermofield_double}).

An object falling into the black hole corresponds to a perturbation that spreads within the circuit \cite{Susskind:2014jwa}. It spreads since the initial perturbation  affects some gates, which then affect some other gates, etc. Eventually all gates are affected by the perturbation. The difference between the time at which the perturbation enters the circuit and the time when all gates are affected is its scrambling time in this quantum circuit.

\begin{figure}[H] 
 \begin{center}                      
      \includegraphics[width=3.6in]{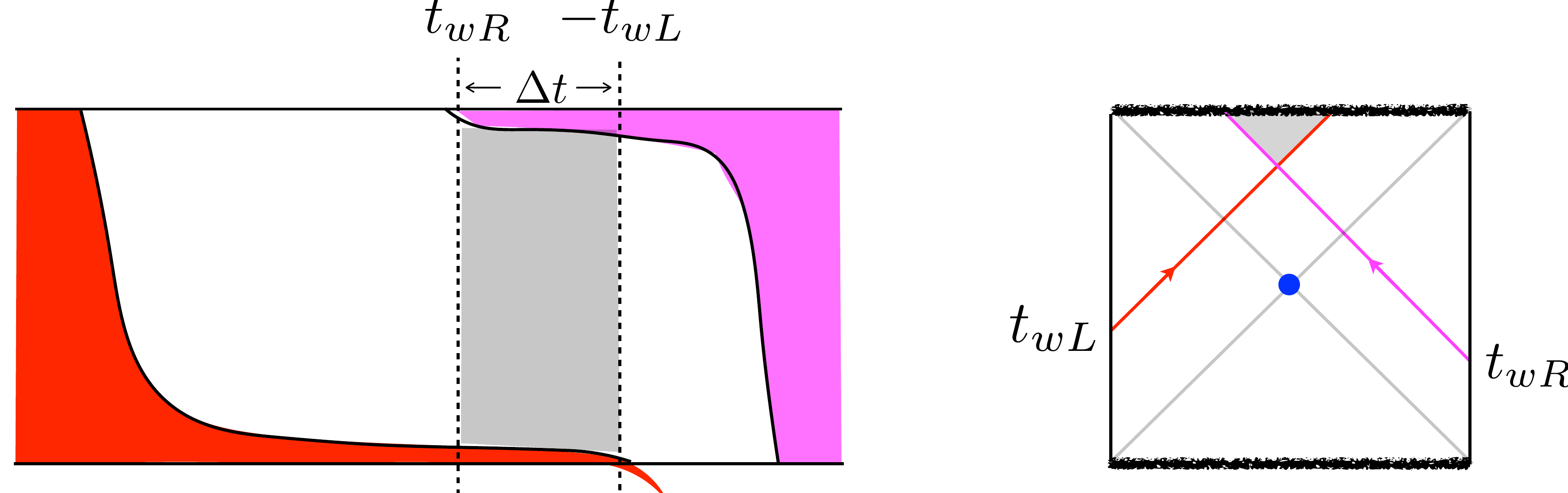}
      \caption{{\it{Left Panel:}} The red perturbation from the left and the pink perturbation from the right have some overlap in the quantum circuit. {\it{Right Panel:}} The two perturbations meet in the black hole interior. It has been proposed that the number of gates in the grey region on the left is related to the spacetime volume of the grey region on the right.}
  \label{overlap_1}
  \end{center}
\end{figure}

 In the case of a Schwarzschild AdS black hole, two infalling objects coming from opposite boundaries may or may not meet in the interior before hitting the singularity. This phenomena has been interpreted as the possible overlap of two perturbations in the shared quantum circuit  \cite{Zhao:2020gxq, Haehl:2021prg}. More precisely, it was proposed in \cite{Zhao:2020gxq} that the number of  gates which are unaffected by either perturbation (called ``healthy gates" in \cite{Zhao:2017isy}) in the overlap region is related to the spacetime volume of the bulk region to the future of the collision between the two objects falling into the black hole \footnote{The intuitive reason is that these gates are applied after both shockwaves come in. So it will correspond to a bulk region which is to the future of both shockwaves.} (see Fig.~\ref{overlap_1}).

\begin{figure}[H] 
 \begin{center}                      
      \includegraphics[width=2.6in]{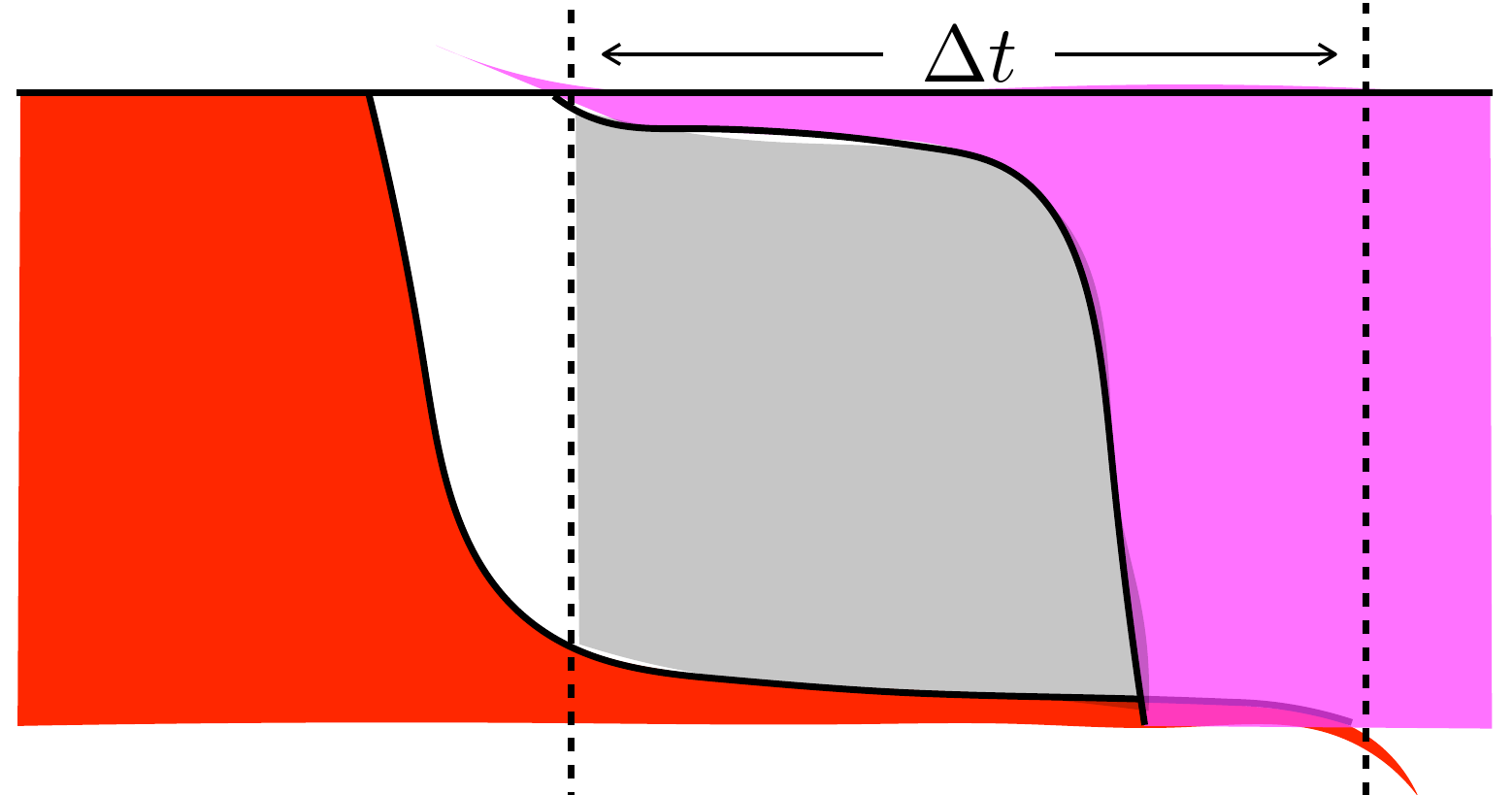}
        \caption{The grey region corresponds to the healthy gates in the overlap, which barely changes as we vary $\Delta t$.}
        \label{plateau_circuit}
  \end{center}
\end{figure}

When the two perturbations overlap and have significantly different scrambling times, the number of gates which are unaffected by either perturbation  remains constant for a while as one varies the time the shells are introduced (see Fig. \ref{plateau_circuit}). In the rest of this section we apply this reasoning to charged black holes. We add either neutral or charged ingoing shells  and calculate how the volume of the
post-collision region changes  when we change the times that the shells are introduced. The existence of a plateau, where the volume remains constant, will be a sign of different scrambling times.

\subsection{Collision between neutral shells with different energy}

\begin{figure}[H] 
 \begin{center}                      
      \includegraphics[width=2in]{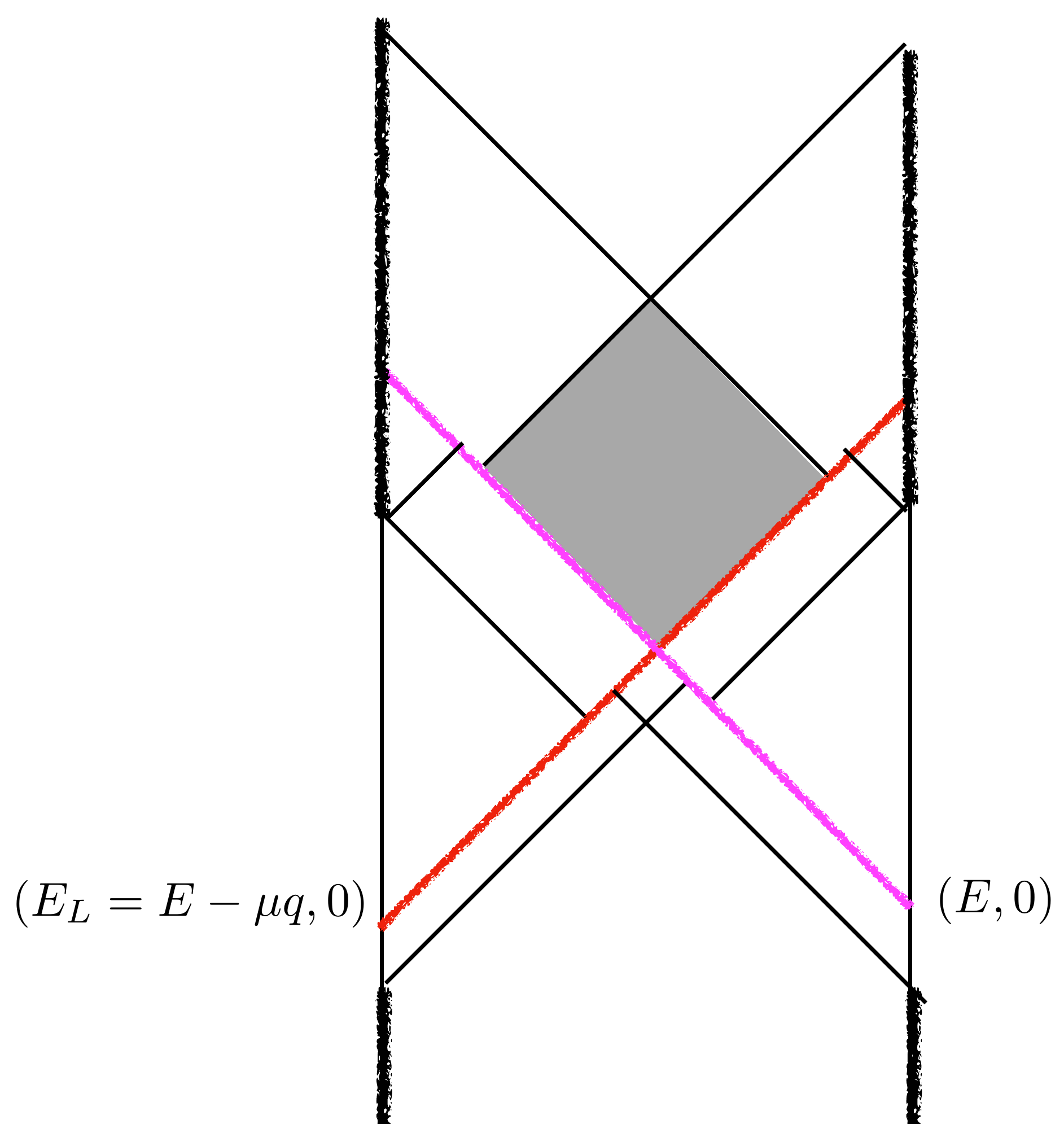}
      \caption{Collision between two neutral shells with different energies. We are interested in the volume of the region to the future of both shocks, up to the inner horizon (shaded in grey).}
        \label{bouncing_1}
  \end{center}
\end{figure}

We first consider the collision of two neutral massless shells thrown into the Reisner-Nordstrom AdS geometry. To mimic the change in $ r_+$ when we later add charge, we will fix an energy $E$ and reduce it by  $\mu q$ for some $q$, where $\mu = Q/r_+$. So one shell carries energy $E_L = E-\mu q$ and comes from the left boundary at time $t_{wL}$, and the other one carries energy $E_R = E$ and comes from the right boundary at time $t_{wR}$ (Figure \ref{bouncing_1}). We consider the regime where $\frac{E-\mu q}{E}\ll 1$ so the scrambling times of the two perturbations are significantly different.

In Appendix \ref{app:post_collision} we show how to construct the spacetime with the two shocks and compute the volume of the region to the future of both shocks. We cut this region off at the inner horizon, since it is unstable and perturbations turn it into a singularity. As mentioned above, from quantum circuit considerations, when the two scrambling times are significantly different, there will be a regime where the number of healthy gates in the overlap region does not change as we vary $\Delta t\equiv -t_{wL}-t_{wR}$. Assuming the same connection between the number of healthy gates and the volume of the post-collision region as discussed above, we expect a  plateau in the post-collision region spacetime volume as a function of $\Delta t$ \cite{Haehl:2022uop}. As we explain in the next section, this is indeed what we find. So for neutral perturbations of a charged black hole, the connection between the number of healthy gates and post-collision volume appears to hold. We are therefore encouraged to extend it to charged perturbations.

\subsection{Collision between a charged shell and a neutral shell}

\begin{figure}[H] 
 \begin{center}                     \includegraphics[width=1.5in]{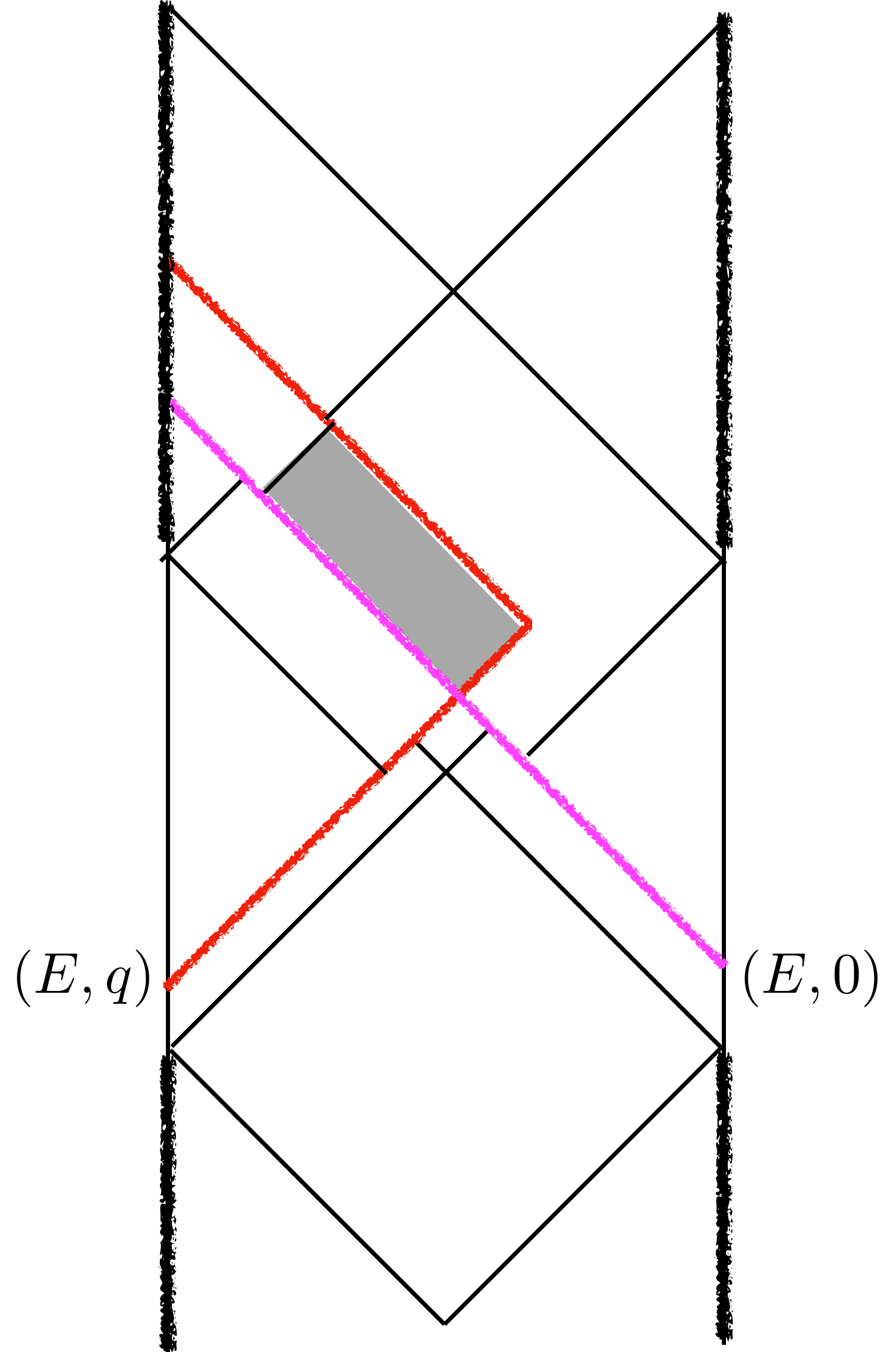}
        \caption{Collision between a charged particle and a neutral particle carrying the same energy. The post-collision region is shaded.}
        \label{bouncing_2}
  \end{center}
\end{figure}

Next, we consider the collision between a charged shell and a neutral shell. We send in a shell with energy $E_L=E$ and charge $q$ from the left boundary at time $t_{wL}$, and another neutral shell with energy $E_R=E$ from the right boundary at time $t_{wR}$. The resulting geometry is shown in Fig. \ref{bouncing_2}. 

\begin{figure}[H]
 \begin{center}                      
      \includegraphics[width=3.7in]{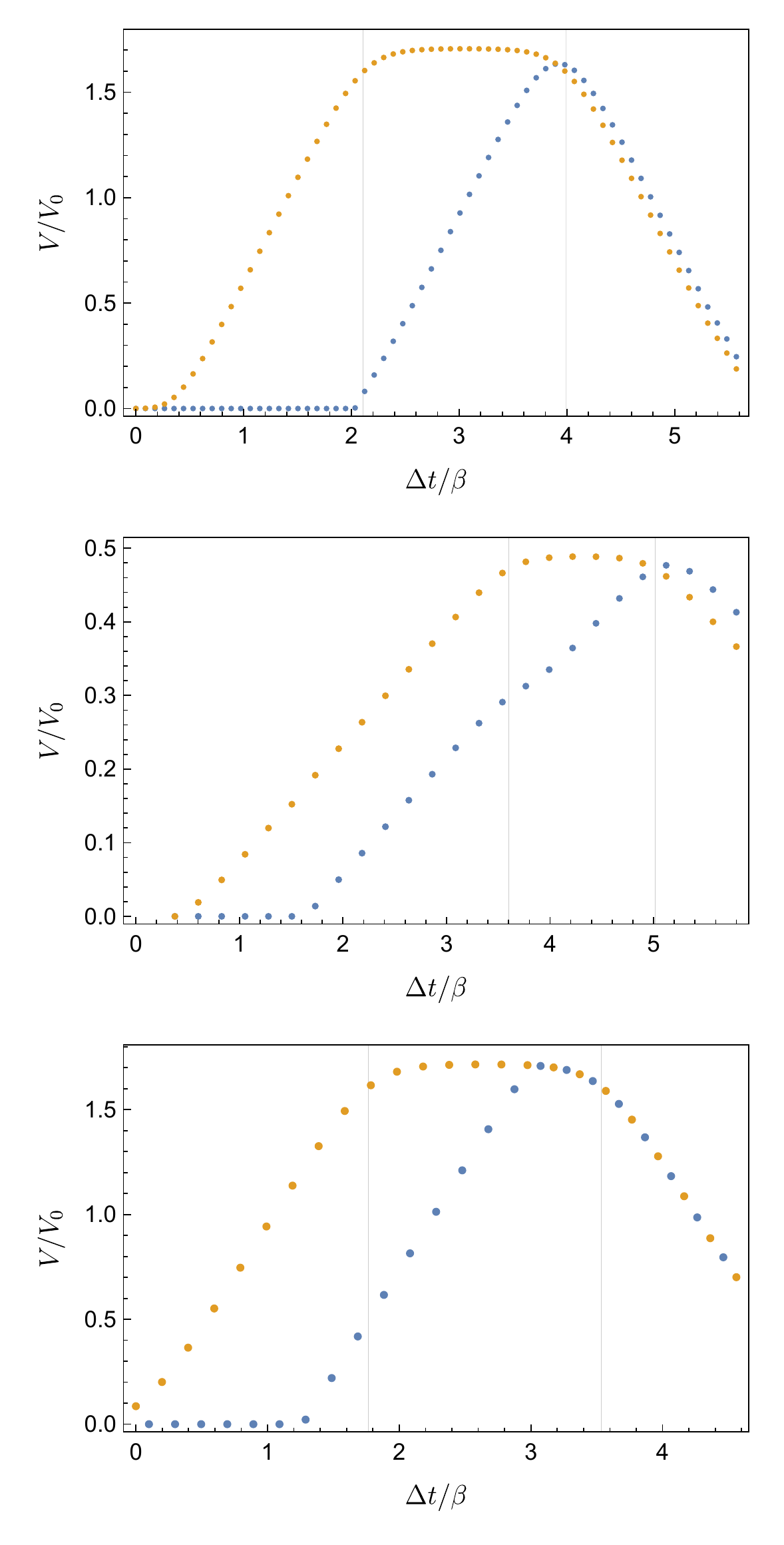}
      \caption{Spacetime volumes of post-collision regions as a function of $\Delta t$ in various examples. The spacetime volumes are given in units of $V_0=4 \pi (r_+^3-r_-^3)\beta/3$ and $\Delta t$ is given in units of $\beta$. The scrambling times of the two neutral shells are marked as vertical grey lines. {\it{Top panel:}} A black hole with medium charge $r_- \sim r_+/2 $ ($r_+/l\simeq 4.82$ and $r_-/l\simeq 2.12$) with perturbations of size $E/M=5\times 10^{-6}$ and $(E-\mu q)/M\simeq 3.74\times 10^{-11}$. {\it{Middle panel:}} A black hole far from extremality ($r_+/l\simeq 31.5$ and $r_-/l\simeq 1.27\times 10^{-4}$) with perturbations of size $E/M\simeq 1.91\times 10^{-9}$ and $(E-\mu q)/M\simeq 2.67\times 10^{-13}$.
      {\it{Bottom panel:}} A black hole close to extremality ($r_+/l\simeq 31.6$ and $r_-/l\simeq 30.5$) with perturbations of size $E/M\simeq 3.33\times 10^{-8}$ and $(E-\mu q)/M\simeq 4.89\times 10^{-13}$.}
      \label{numericall}
  \end{center}
\end{figure}

The construction of the geometry and calculation of the volume of the post-collision region in this case is more involved, and is explained in Appendix \ref{app:post_collision}. We have numerically computed this volume
as a function of $\Delta t = -t_{wL}-t_{wR}$ for several choices of mass $M$ and charge $Q$ of the original black hole. In each case, we choose $E\ll M$ and $q\ll Q$ so that $E-qQ/r_+ \ll E$ . The final results are shown by the blue dots in Fig. \ref{numericall}. For comparison,  the results for the case the two neutral shells (discussed in the previous subsection) are shown by the orange dots.

The top plot in Fig. \ref{numericall} is for an average charged black hole (with $r_- \sim r_+/2) $. The next one has very little charge and the bottom plot is close to extremality. In all cases,  one can see that the orange dots have a plateau in the middle as expected from the quantum circuit analysis, which is a signal that the two scrambling times are significantly different. For the blue dots,  instead of a plateau in the middle, in all cases one sees a delay at the beginning. In the first two plots, the initial delay, $t_d$, equals the width of the plateau which is the result of the coincidence we discussed in Sec. 2.  This suggests that away from extremality, the scrambling of the charged particle takes the same amount of time as the scrambling of the neutral particle with the same energy, but the onset of the process, i.e., the time at which the perturbation enters the circuit, is delayed by $t_d$. 

There is a simple explanation for this difference in the behavior of the post-collision volume, between charged and neutral shells.  First consider the neutral case represented by the orange dots. When $\Delta t$ is small, the two shells collide with small center of mass energy. So the backreaction is negligible and the volume can be computed in RN AdS, and shown to increase linearly.\footnote{The post-collision volume at $\Delta t =0$ is nonzero, but it is small compared to $V_0  =4 \pi (r_+^3-r_-^3)\beta/3$  which is a convenient scale to measure the volume in.}
As $\Delta t$ increases, the shells are introduced earlier, and their center of mass energy increases. Eventually, the backreaction becomes important, causing the black hole to grow. This is the start of the plateau. While it is clear that the linear growth of the post-collision volume will be modified, it is rather surprising from this perspective that the volume remains constant.  At very large center of mass energy, the collision takes place deep inside a large black hole and the post-collision volume up to the inner horizon goes to zero.

Now consider the case of a charged perturbation. Since a charged shell bounces, there is no collision  unless the shell on the right starts before a time $t_b$ (see Fig. \ref{trajectory_bouncing}). So the post-collision volume is strictly zero until $\Delta t  = -t_{wL} - t_b \equiv t_d$. After this, the volume grows linearly for a while (for the same reason as the neutral case), and starts to decrease  when the backreaction becomes important.

\section{Probing the bounce in a two-dimensional model}
\label{sec:interior_size}
To gain more insight into the effects of a charged perturbation bouncing inside the horizon, in this section we consider a  simple two-dimensional model. We will see that the bounce can be detected by the 
following out-of-time-ordered  four-point correlator (OTOC)

 \begin{align}\label{OTOC}
    \langle V_L (t_L) V_R(t_R) \rangle_W \equiv \frac{\bra{TFD}W^{\dagger}(t_{wL})V_L(t_L)V_R(t_R)W(t_{wL})\ket{TFD}}{\bra{TFD}W^{\dagger}(t_{wL})W(t_{wL})\ket{TFD}}\,,
 \end{align}
 for a judicious choice of  times for the operator insertions. Here 
 $W$ corresponds to a charged perturbation and $V$ is a neutral operator.\footnote{It is equally interesting to consider $V$ to be a charged operator. However, for the sake of simplicity, we choose to focus on the neutral case in what follows.}
 We will distinguish two choices of times. If  $t_R = -t_L$, we obtain the standard OTOC which has been used to discuss chaos and the onset of scrambling \cite{Maldacena:2015waa}. We will call this the ``exterior OTOC" since it only probes the perturbation outside the horizon of the original black hole.  If instead we take $t_L $  very large, we obtain something we will call the ``interior OTOC", since we will see it probes the perturbation inside the black hole.
 This OTOC not only can detect the bounce, but also gives another indication of a scrambling delay.

\subsection{Bounces in AdS$_2$}
When focusing on operators $V$ with large scaling dimensions $\Delta$ one can sometimes exploit the WKB approximation to write correlators in terms of geodesic lengths \cite{Louko:2000tp}
\begin{equation} \label{geodesic_aprox}
\langle V_L (t_L) V_R(t_R) \rangle_W \sim e^{-\Delta L(t_L, t_R)}\,,
\end{equation}
where $L(t_L, t_R)$ corresponds to the appropriately renormalized geodesic distance connecting both boundaries at the chosen times in the geometry perturbed by $W$. We will use this to compute  \eqref{OTOC}.

In particular, we will consider a perturbation $W$ described by charged particles with a turning point in an AdS$_2$ geometry. Such particles arise in the Jackiw-Teitelboim (JT) gravity obtained through the dimensional reduction of a near extremal electrically charged black hole \cite{Moitra:2018jqs,Goto:2018iay}, where we end up with an AdS$_2$ geometry equipped with an electric field and a dilaton $\Phi$. For the sake of simplicity, we will be agnostic regarding the particular theory in which our setup is embedded and focus only on having the metric and the dilaton be $C^0$ in an appropriate coordinate system. We will write the AdS$_2$ geometry in the following coordinates 
\begin{equation}
ds^2=-f(r)dt^2+f(r)^{-1}dr^2
\end{equation}
with 
\begin{equation}
f(r)=(r-r_{+})(r-r_{-})\,,
\end{equation}
where we are setting the AdS length equal to one. We will assume that the dilaton takes the form 
\begin{equation}
\Phi=\Phi_h r\,
\end{equation}
in these coordinates, where $\Phi_h$ is some arbitrary constant. This assumption is motivated by the fact that in the context of a spherically symmetric dimensional reduction the dilaton captures the size of the higher dimensional sphere. More concretely, this happens in the JT-like gravity described in \cite{Goto:2018iay} after an appropriate coordinate redefinition. The Kruskal null coordinates are still defined as in eqs. \eqref{Kruskal_1} and \eqref{Kruskal_2} but now we can explicitly find the tortoise coordinate in eq. \eqref{tortoise}, which yields

\begin{equation}
r^*(r)=-\int_r ^\infty \frac{dr'}{f(r')}=\frac{\beta}{4\pi}\log{\left(\frac{r-r_{+}}{r-r_{-}}\right)}\,,
\end{equation}
with
\begin{equation}
\beta=\frac{4\pi}{f'(r_{+})}=\frac{4\pi}{r_{+}-r_{-}}\,.
\end{equation}

Let us now consider that we send a massless particle (which might carry electric charge) from the left boundary of the AdS$_2$ black hole at some time $t_{wL}$. This particle will initially follow a trajectory of constant $U_w=e^{2\pi t_{wL}/ \beta}$. We will assume that the `backreaction'\footnote{For details on the exact meaning of backreaction in this situation, see Appendix B in \cite{Maldacena:2016upp}. An alternative way to think about it is that the two-dimensional model can represent the near horizon region of a higher dimensional near extremal black hole.
A perturbation in higher dimensions changes the parameters of the black hole. As a consequence, a stationary observer measures a slightly different temperature before and after the shell. A natural way to model this effect in the approximate AdS$_2$ near horizon geometry is to change the Rindler patch.} of this particle is non-negligible and thus we must take a different Rindler patch of AdS$_2$ after the particle is sent in, which we will denote by
\begin{equation}
\tilde{ds}^2=-\tilde{f}(r)d\tilde{t}^2+\tilde{f}(r)^{-1}dr^2
\end{equation}
with 
\begin{equation}
\tilde{f}(r)=(r-\tilde{r}_{+})(r-\tilde{r}_{-})\,.
\end{equation}
The coordinate $r$ is taken to be continuous across the trajectory of the particle in order to impose continuity of the dilaton. As before, it is natural to take the time coordinate to flow continuously along the boundary, which implies that the shell follows a trajectory of constant $\tilde{U}_w=e^{2\pi t_{wL}/ \tilde{\beta}}$ in the new coordinates. Continuity of the $r$ coordinate implies the following relationship along the constant $U$ trajectory of the shell
\begin{equation}\label{VVtilde}
\tilde{V}=\frac{1}{\tilde{U}_w}\frac{r_{+}+r_{-}U_w V-\tilde{r}_+(1+U_w V)}{\tilde{r}_{-}-r_{+}+U_w V(\tilde{r}_{-}-r_{-})}
\end{equation}
If the particle does not bounce, this fully describes the trajectory of the shell in both patches. We are interested in considering shells that do bounce at some {radius} $r_b$. If this happens, in analogy to higher dimensions, the DTR condition \cite{Dray:1985yt, 10.1143/PTP.73.1401, PhysRevD.41.1796} should be satisfied at the turning point in order for the metric to be $C^0$ in appropriate coordinates. In this case, it implies 
\begin{equation}\label{r_b_eq}
f(r_b)=\tilde{f}(r_b) \Longrightarrow r_b=\frac{r_{-}r_{+}-\tilde{r}_{-}\tilde{r}_{+}}{r_{+}+r_{-}-\tilde{r}_{+}-\tilde{r}_{-}}\,.
\end{equation}
If the shell does bounce at some $r_b$, it will start following a trajectory of constant $V =V_b$  with
\begin{equation}\label{V_b}
V_b=\frac{1}{U_w}\frac{r_{+}-r_b}{r_b-r_{-}}
\end{equation}
and analogously for $\tilde V =\tilde{V}_b$ with appropriate replacement by tilded quantities. Again, using the continuity of $r$, we can find how $\tilde{U}$ and $U$ relate to each other along the shell after the bounce
\begin{equation}\label{UUtilde}
\tilde{U}=\frac{1}{\tilde{V}_b}\frac{r_{+}+r_{-}U V_b-\tilde{r}_+(1+U V_b)}{\tilde{r}_{-}-r_{+}+U V_b(\tilde{r}_{-}-r_{-})}\,.
\end{equation}

 \subsection{OTOC Behavior}
With the setup described above in mind, one can compute the geodesic length between both boundaries at times $t_L$ and $t_R$ with the use of embedding coordinates. This computation is carried out in detail in Appendix \ref{app:ads2_geodesic} and yields
\begin{equation}\label{geodesic_distance}
\begin{split}
e^{-L}&=\frac{64\pi^4 e^{\Delta t_L}e^{\Delta t_R}}{\beta^2 \tilde{\beta}^2 r_c^2 \left[\delta r_{-}+(\tilde{r}_{-}-r_{+})e^{\Delta t_R}+(r_{-}-\tilde{r}_{+})e^{\Delta t_L}-\delta r_{+}e^{\Delta t_L}e^{\Delta t_R}\right]^2}\,,
\end{split}
\end{equation}
where we defined 
\begin{equation}
\delta r_{+}=\tilde{r}_{+}-r_{+}\,,\,\,\, \delta r_{-}=\tilde{r}_{-}-r_{-}\,
\end{equation}
and\footnote{Note that in this section we define $\Delta t$ to be dimensionless.} 
\begin{equation}\label{time_difference}
\Delta t_{L}=\frac{2\pi}{\tilde{\beta}}(t_{L}-t_{wL})\,,\,\,\, \Delta t_{R}=\frac{2\pi}{\beta}(-t_{wL}-t_R)\,.
\end{equation}
The variable $r_c$ corresponds to the radial cut-off. It is important to keep in mind that this expression is only valid for $\Delta t_L>0$, i.e., when evaluating the geodesic length after the insertion of the operator $W$. In general, the insertion of the operator should be thought of as creating both an ingoing and an outgoing particle and this expression does not take that into account. 

By studying eq. \eqref{geodesic_distance} as a function of $t_{wL}$ for different choices of perturbations and times $(t_L,t_R)$ one can notice an interesting feature. When the particle does not bounce between the two horizons, the OTOC monotonically decreases as $t_{wL}$ is moved into the past. This is qualitatively the same as the standard result found in \cite{Shenker:2013pqa}. However, when the bounce happens in between the horizons, the OTOC can have a maximum at some $t_{wL}^{*}<t_L$ and only monotonically decays for $t_{wL}<t_{wL}^{*}$ as we move the perturbation further into the past. As we will show in more detail below, this maximum occurs exactly when the geodesic between the two sides passes through the turning point of the particle and thus it provides a boundary probe of the bounce inside the black hole interior. Naturally, this only happens for suitable choices of times $(t_L,t_R)$ such that the geodesic meets the particle in the interior.

To study this behavior in more detail we turn to a particular derivative of the geodesic length, namely
\begin{equation}\label{length_derivative}
-\frac{d}{dt_{wL}}L \approx \frac{2\pi}{\beta}\left(\frac{(\delta r_{-}-\delta r_{+}) e^{\Delta t_L}+(\delta r_{+}-\delta r_{-})e^{\Delta t_R} -2\delta r_{-} -2\delta r_{+} e^{\Delta t_L}e^{\Delta t_R}}{(r_{-}-r_{+})e^{\Delta t_R}+\delta r_{-}+(r_{-}-r_{+}) e^{\Delta t_L}-\delta r_{+}e^{\Delta t_L}e^{\Delta t_R}}\right) \,,
\end{equation}

where we assumed $r_{+}-r_{-} \gg \delta r_{\pm}$ and worked to leading order \footnote{It is important to only do this approximation after taking the derivative as otherwise one will neglect contributions which are non-negligible for certain time configurations. In particular, when doing this approximation one should be careful to account for the fact that small terms might be enhanced by the exponentials.}. This can be achieved by having the perturbation be much smaller than the background and, at the same time, by not being too close to extremality. This derivative removes the renormalization ambiguity $r_c$, and its vanishing will single out when the maximum of the OTOC occurs. It is perhaps not surprising that this maximum occurs at the turning point of the particle. It was shown in \cite{Lin:2019qwu} that for neutral particles in two dimensions, precisely this derivative of the length gives the momentum of the particle along the geodesic.  Even though our particle is charged, it still captures the fact that the momentum vanishes at the turning point.

\subsubsection*{Exterior OTOC} 

Having chosen $t_{wL}$, the time at which the boundary operator $W$ is inserted, we can still choose the boundary times $(t_L,t_R)$ at which we wish to insert the probe operators $V$. 

\begin{figure}[H] 
\begin{center}                     \includegraphics[width=1.5in]{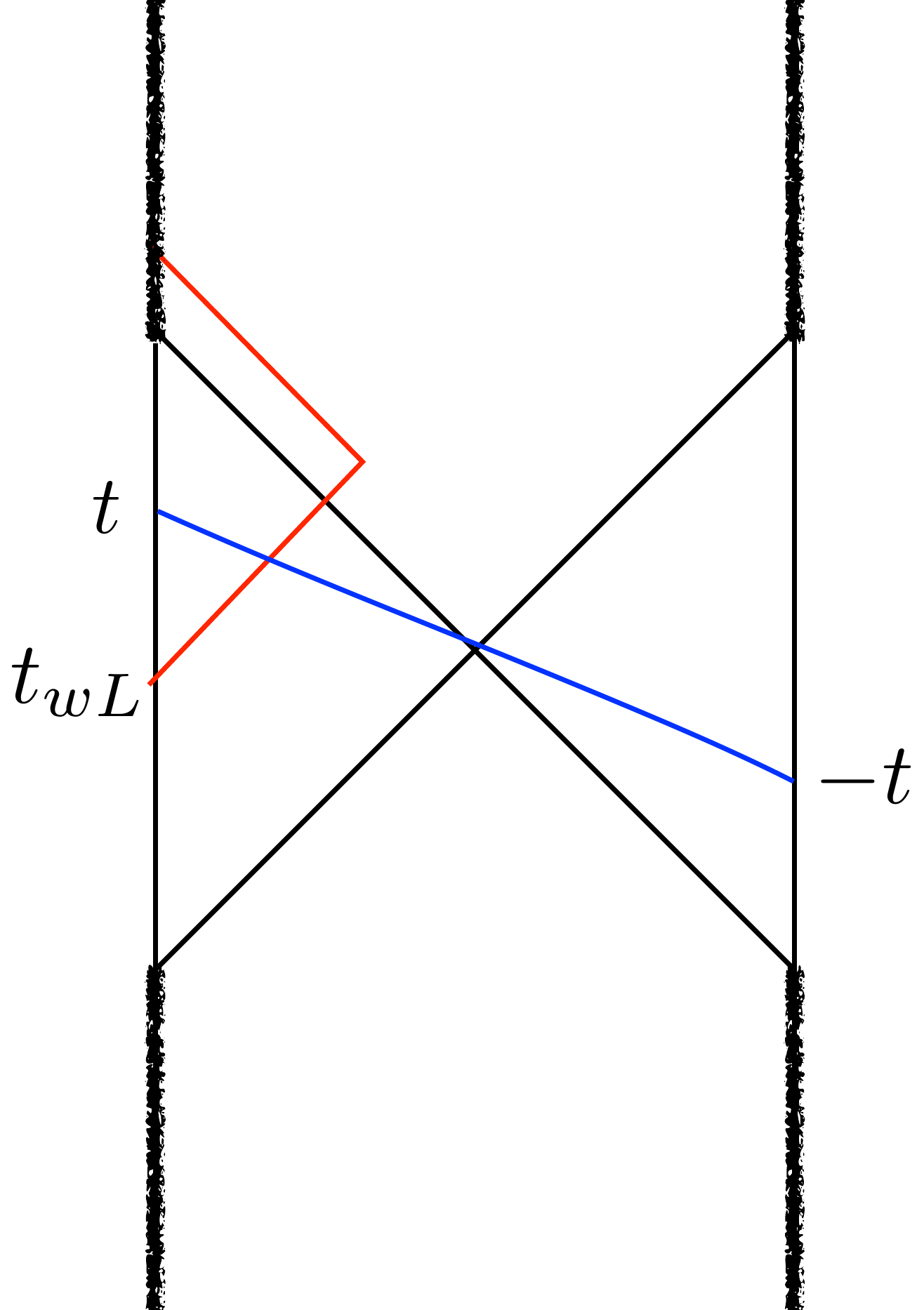}
        \caption{Exterior OTOC for the AdS$_2$ black hole}
        \label{correlator_1}
\end{center}
\end{figure}

We will start by considering the standard choice made in \cite{Shenker:2013pqa} by setting $t_L = -t_R = t$.\footnote{As mentioned at the end of Section \ref{sec:bounce_inside_bulk}, Ref. \cite{Shenker:2013pqa} considered the case $t=0$, but due to the timelike Killing symmetry this is equivalent to keeping $t$ free. This is made manifest by the fact that the final result \eqref{size_exterior} depends only on the the difference $\Delta t$.} 
%This choice defines a specific slicing of the black hole - it is natural to think of it as an `exterior slicing' 
We call this the ``exterior OTOC" since the geodesic connecting the two boundaries at these times does not pass through the black hole interior in the unperturbed geometry. Moreover, using the results in Appendix \ref{app:ads2_geodesic}, one can explicitly check that the geodesic always meets the particle outside the initial black hole. Therefore this choice cannot directly probe the presence of a bounce inside  (see Fig. \ref{correlator_1}).

\noindent With the aforementioned choice of times, using eq. \eqref{length_derivative}, we find 
\begin{align}
\label{size_exterior}
	&-\frac{d}{dt_{wL}}L\Bigg|_{t_L = -t_R = t} \approx \frac{4\pi}{\beta}\left(\frac{ \delta r_+ e^{\Delta t}+ \delta r_- e^{-\Delta t}}{2(r_{+}-r_{-})+\delta r_+e^{\Delta t}- \delta r_- e^{-\Delta t}}\right)\,,
\end{align}
where we defined 
\begin{equation}\label{eq:deltat}
\Delta t= \frac{2\pi}{\beta}(t-t_{wL})\,.
\end{equation}

As we remarked before, our expressions are only valid for $\Delta t_L>0$ which in this case amounts to $\Delta t>0$. One can quickly check that eq. \eqref{size_exterior} is always positive for $\Delta t>0$. Furthermore, it grows as $\Delta t \to \infty$, saturating at a value of $4\pi/ \beta$.\footnote{Had we written a valid expression for all $\Delta t$, we would naturally find it to be anti-symmetric under $\Delta t\rightarrow -\Delta t$
\begin{equation}\label{anti_sym_size}
    -\frac{d}{dt_{wL}}L\Bigg|_{t_L = -t_R = t} \approx \frac{4\pi}{\beta}\left(\frac{ \delta r_+ e^{|\Delta t|}+\delta r_- e^{-|\Delta t|}}{2(r_{+}-r_{-})+ \delta r_+e^{|\Delta t|}- \delta r_- e^{-|\Delta t|}}\right)\text{sgn}(\Delta t)\,.
\end{equation}
From one perspective, this reflects the fact that the length itself should be invariant under this transformation if the insertion of the operator simultaneously creates an outgoing and an ingoing particle. Alternatively, this derivative is expected to capture an appropriate notion of momentum and therefore this anti-symmetry can be seen as a natural consequence of the fact that each particle carries the same momentum in absolute value, although one is ingoing and the other is outgoing.}   This implies that, regardless of the presence of a bounce in the interior, the OTOC obtained through this choice of boundary times starts decreasing immediately at $\Delta t=0$ and monotonically decays as a function of $\Delta t$. This is no different from the standard result obtained in \cite{Shenker:2013pqa}. 

There is a sense, however, in which the bounce does affect this OTOC. Using eq. \eqref{r_b_eq}, we can write 
\begin{equation}\label{deltarp_as_rb}
\delta r_{+}=\frac{\delta r_{-}(r_{+}-r_{b})}{r_b-r_{-}-\delta r_{-}}
\end{equation}
from which we can see that  $r_{b}\to r_{+}$ implies $\delta r_{+} \to 0$. It follows that as the bounce happens closer to the horizon, the derivative computed above takes a longer time to become $\mathcal{O}(1)$, essentially because the growth is controlled by the term $\alpha= \delta r_+ e^{\Delta t}$. We have called it $\alpha$ for a reason. This is in every way analogous to the behavior of the parameter $\alpha$ we described in Section \ref{sec:bounce_inside_bulk} where a bounce closer and closer to the horizon implies, for a fixed $t_{wL}$, a smaller and smaller $\alpha$ which vanishes in the limit, implying what appeared to be a longer scrambling time. This is not surprising. After all, while we did not frame this discussion in terms of $\alpha$, the OTOC we just considered is computed through a spacelike probe at $t_L=-t_R=t$ which is exactly the kind of probe affected by $\alpha$. %An important difference is that in higher dimensions the relationship between such correlators and geodesics is more subtle to establish \cite{Reynolds:2016pmi,Fidkowski:2003nf}. 
We should remember, however, that as emphasized previously this effect cannot distinguish a shell with charge from a shell without charge but less energy. While in our rough toy model we have not introduced  notions of energy or charge, we can observe that in the limit $r_b \to -\infty$, in which the particle never bounces, we have $\delta r_{+} =-\delta r_{-}$ and so we can make $\delta r_{+}$ as small as we wish by decreasing $\delta r_{-}$ in absolute value\footnote{We must have $\delta r_{-}<0$ in this case in order to preserve the second law: $\delta r_{+}>0$.}. This is equivalent to the fact that we can decrease $\delta r_{+}$ in higher dimensions by simply decreasing the energy of the shell instead of making it bounce closer to the horizon by adding charge - in fact, the two statements are the same  if we word everything in terms of $\delta r_{-}$ and $\delta r_{+}$ instead of $E$ and $q$. 
However, in contrast to the discussion in Section \ref{sec:bounce_inside_bulk}, we are not taking a large $\Delta t$ limit \footnote{See \cite{Caputa:2015waa} for an analysis of scrambling in BTZ which generalizes the results in \cite{Shenker:2013pqa} to finite $\Delta t$.}. If we did so, $\delta r_{-}$ contributions would become negligible and \eqref{size_exterior} would depend only on $\alpha$.  Since at finite $\Delta t$ the contributions arising from $\delta r_{+}$ and $\delta r_{-}$ can be comparable, even if small, \eqref{size_exterior} encodes some distinguishable, although indirect, effects of the bounce. This follows from the fact that $r_b$ is encoded in the relationship between $\delta r_{+}$ and $\delta r_{-}$ through \eqref{deltarp_as_rb}.

The bottom line of the above discussion is that despite the fact that \eqref{size_exterior} is indirectly affected by the bounce at finite $\Delta t$, the bounce does not change its qualitative behavior. In particular, the derivative is always positive and thus the OTOC always decreases as a function of $\Delta t$.

\subsubsection*{Interior OTOC}
\begin{figure}[H] 
 \begin{center}                     \includegraphics[width=1.5in]{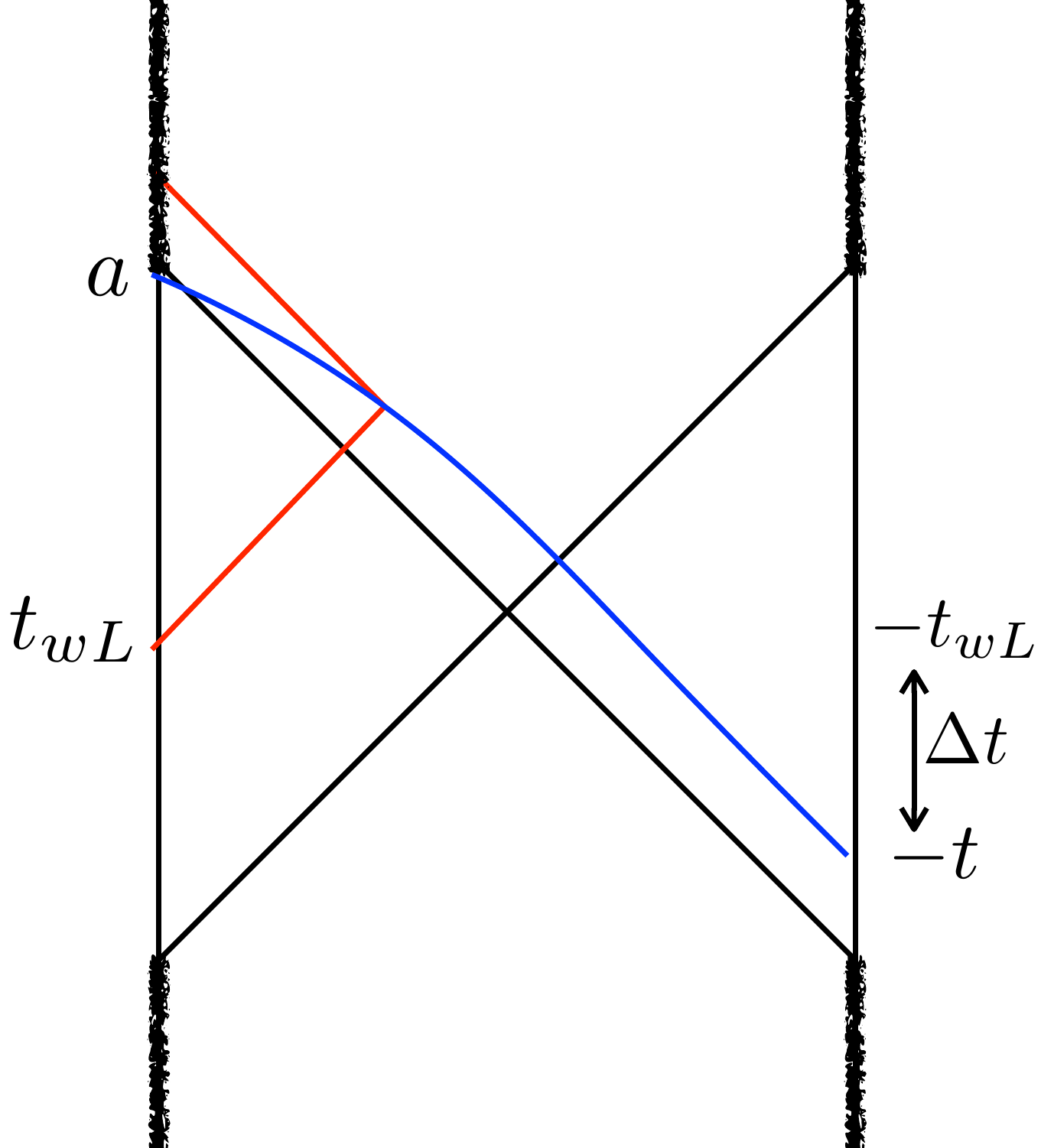}
        \caption{Interior OTOC for the AdS$_2$ black hole}
         \label{correlator_2}
  \end{center}
\end{figure}

We will now focus on a different choice of times for the operator insertions so the OTOC will probe the interior of the black hole. Namely, we take $(t_L, t_R) = (a,-t)$ where $a$ is positive and larger than the scrambling time of the perturbation.  
%In this limit, 
%different choices of $t_R$ provide a slicing of the black hole interior as illustrated in Figure \ref{correlator_2}. 
Contrary to the choice studied before, this one ensures that the geodesic connecting both sides meets the particle in the interior of the black hole (see  Fig. \ref{correlator_2}), and therefore we expect it to directly capture the presence of a turning point in its trajectory.

\noindent For simplicity, we will focus on the limit $a \to + \infty$ of the derivative \footnote{It is important to keep in mind that the OTOC itself will vanish in this strict limit. We take this limit because it approximates the behavior of the derivative when $a$ is large but finite by a simpler expression where the $\Delta t_L$ dependence drops out.}

\begin{align}
\label{size_interior}
	&-\frac{d}{dt_{wL}}L\Bigg|_{ t_L\rightarrow \infty} \approx \frac{4\pi}{\beta}\left(\frac{ \delta r_+ e^{\Delta t}-(\delta r_--\delta r_+)/2}{(r_+-r_-)+  \delta r_+ e^{\Delta t}}\right)
\end{align}
Since we have ensured that $\Delta t_L$ is always positive, this expression is valid for any finite $\Delta t$.

\begin{figure}[H] 
 \begin{center}                     \includegraphics[scale=1.2]{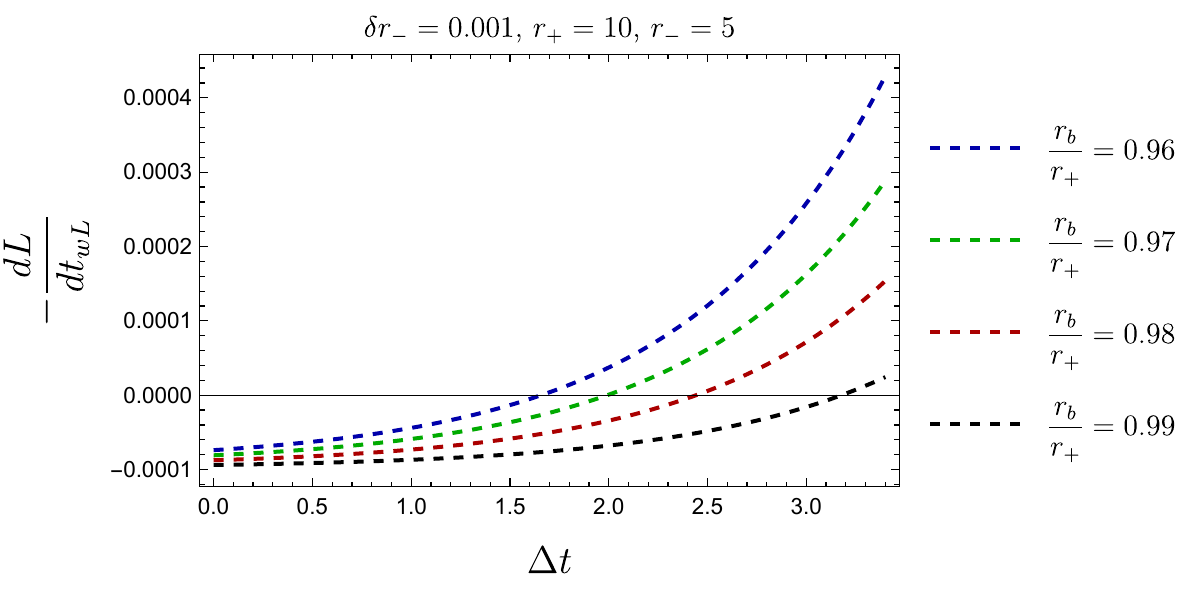}
        \caption{Plot of the derivative of the length associated to the interior OTOC for different particles. Notice how the zero shifts towards larger $\Delta t$ as the particle bounces closer to the horizon.}
         \label{interior_derivative}
  \end{center}
\end{figure}

While, just like eq. \eqref{anti_sym_size}, the growth of this expression is dominated by $ \delta r_{+}e^{\Delta t}$ and saturates at the same value $4\pi/\beta$, we see that, contrary to the former (having in mind \eqref{anti_sym_size}), a significant growth only occurs in the direction  of positive $\Delta t$. Moreover, it vanishes at
\begin{equation}\label{t_delay2}
\Delta t=t_d^* \equiv \log{\left(\frac{\delta r_{-}-\delta r_{+}}{2\delta r_{+}}\right)} \approx \log{\left(\frac{r_b-\frac{r_{-}+r_{+}}{2}}{r_{+}-r_b}\right)}\,,
\end{equation}
where we used eq. \eqref{deltarp_as_rb} to write the last equality while working to leading order in the perturbation. This zero exists provided that $\delta r_{-}>\delta r_{+}$ or in terms of the position where the bounce occurs $r_b > (r_{-}+r_{+})/2$ \footnote{We remind the reader we are always assuming bounces inside the horizon, i.e. $r_b<r_{+}$.}, i.e. when the bounce occurs closer to the outer horizon than to the inner horizon. This means that if a bounce happens too far inside the black hole $r_b <
 (r_{-}+r_{+})/2$, eq. \eqref{size_interior} is always positive. Notice that the vanishing of the derivative occurs in a regime in which all $\delta r_{+}$ and $\delta r_{-}$ contributions are comparable.  We illustrate this feature in Fig.  \ref{interior_derivative}.
 
\noindent This zero in the derivative can be shown to imply the existence of a local maximum in the OTOC as a function of $t_{wL}$.  We now show that this maximum occurs exactly when the geodesic meets the particle at the bounce.   Thus the existence of a maximum can be seen as a direct boundary probe of a bounce in the interior.  In Appendix \ref{app:ads2_geodesic}, we derived $V_m$ \eqref{V_meet}, the value of the coordinate $V$ at which the geodesic meets the particle along its constant $U$ part of the trajectory. In the limit $t_L \to + \infty$, we can ask at which $\Delta t$ we find $V_m=V_b$ \footnote{In the limit $t_L \to + \infty$, the time dependence for this equation comes only through $\Delta t$.}, where $V_b$ corresponds to the value of the coordinate at which the particle bounces and is given in eq \eqref{V_b}. We find the only solution of this equation to be $\Delta t=t_d^*$, i.e. the same time difference at which the derivative vanishes. The reason why the zero in the derivative disappears for bounces too far inside the black hole can then be easily understood from the fact that the geodesics connecting both sides do not reach arbitrarily close to the inner horizon. 

In summary, we see that the interior OTOC is particularly sensitive to bounces close to the horizon at finite $\Delta t$. In particular, the onset of its monotonic decay is delayed.  At this point, one could worry about the non-uniqueness of our choice of OTOC. After all, a priori, despite the clear connection between this maximum and the bounce, a different choice of times for the operator insertions could also have maxima as a function of $t_{wL}$ even without a bounce in the interior, making the presence of a maximum less reliable as a marker of an interior bounce. We now show this cannot happen. Equating \eqref{length_derivative} to zero, we find
\begin{equation}
r_b=\frac{1}{2}\left(r_{-}+r_{+}+\frac{r_{+}-r_{-}}{-1+e^{\Delta t_L}}+\frac{r_{+}-r_{-}}{1+e^{-\Delta t_R}}\right)
\end{equation}
where we used \eqref{deltarp_as_rb} to write $\delta r_{+}$ as a function of $r_b$. Since $\Delta t_L>0$, we see that eq. \eqref{length_derivative} can only vanish if $r_b>(r_{+}+r_{-})/2$. In other words, there can only be extrema in the OTOC as a function of $t_{wL}$ if a bounce occurs closer to the outer horizon than to the inner one. This establishes, without loss of generality, an intimate connection between the bounce in the interior and the presence of extrema in the OTOC.\footnote{We say extrema rather than maxima here since for $t_{wL}$ close to $t_{L}$, the OTOC can have a local minimum, as can be checked by equating \eqref{length_derivative} to zero. This minimum is lost in eq. \eqref{size_interior} due to the limit we are taking.}

\subsection{Remarks on scrambling delay}

We saw above that the interior OTOC has the property that the onset of its monotonic decay is
 delayed due to the presence of a bounce in the interior. If we define the onset of scrambling as the moment at which the OTOC starts to monotonically decrease, it is natural to relate this behavior to the scrambling delay discussed in Section \ref{sec:collision}. In fact, our choice of the variable $t_d^*$ in \eqref{t_delay2} was not arbitrary. In the context of Reissner-Nordstrom-AdS, we defined $t_d$ in \eqref{t_delay1} by considering the latest time that an observer from the right can jump in and still  meet the bouncing shell. We then argued that $t_d$  captures, in some sense, a delay in scrambling. If we do the same thought experiment in our background AdS$_2$ geometry, we find the analogous time \footnote{Again, we are defining a dimensionless $t_d$, contrary to what we did in the previous sections.}
\begin{equation}
t_d= \log{\left( \frac{r_{-}-r_b}{r_b-r_{+}}\right)} \approx \log{\left( \frac{\delta r_{-}}{\delta r_{+}}\right)}\,.
\end{equation}
Comparing this result with eq.\eqref{t_delay2}, we see that when the bounce happens close to the horizon $r_b \to r_{+}$, which is equivalent to $\delta r_{-} \gg \delta r_{+}$, we have $t_d=t_d^* + \mathcal{O}(1)$, i.e. this time difference agrees with the amount of delay for the onset of decay of the interior OTOC up to an order one correction. Just like before, we can only talk concisely about a scrambling delay in the limit in which we have a scale separation between the amount of delay and the thermal time $\beta$. This agreement further supports the statement that the interior OTOC, similarly to the  quantum circuit, is capturing a delay in scrambling due to the interior bounce. It is important to emphasize that this effect is present only for particular choices of OTOC. The onset of scrambling is not universally delayed, but probes of the black hole interior which are blind to the exterior physics appear to take longer to notice scrambling is occurring.

\section{Discussion}
\label{sec:discussion}

Massless charged shells falling into a charged black hole bounce at a particular radius. Starting with an asymptotically AdS black hole, we have investigated what effect this has for scrambling in the dual field theory. We saw that a standard shock wave calculation of a scrambling time indicates that adding charge to a perturbation at early time increases the scrambling time. We then gave evidence that, at least from a certain viewpoint, scrambling does not take longer. Instead, the onset of scrambing is delayed relative to a neutral perturbation with the same energy. 

The evidence consisted of two calculations. The first starts with two shells, one neutral and one charged, that collide inside the black hole. Arguments based on quantum circuits have suggested that the volume of the post-collision region can diagnose whether the scrambling times of the dual neutral and charged perturbations are the same or very different. We computed this volume and found that it behaves as if the scrambling times are the same, but the charged scrambling is delayed. 
We next computed an OTOC in a simple two-dimensional model.  With one choice of times for the operator insertion we get a standard exterior OTOC which probes the perturbation outside the black hole. With another choice of times, we get an interior OTOC which is sensitive to the perturbation inside the horizon. As we saw above, this interior OTOC detects the bounce of the charged particle, and only starts to decay after a certain delay that agrees with the delay calculated earlier.

There are many open questions, and we mention just a few. First, our calculations mainly probe the case when the charged shell bounces inside, but close to the horizon.  (The connection between the quantum circuit and post-collision volume is believed to hold only for collisions and bounces near the horizon.) It would be interesting to determine what are the effects of a bounce farther inside the black hole.
Next, it is clearly of great interest to have a more microscopic understanding of why scrambling is delayed for certain charged perturbations. This can perhaps be studied in the context of a charged SYK model \cite{Gu:2019jub}.
Besides that, our calculation of the OTOC is based on two-sided eternal black holes, but we expect the lesson we learned to be general. For one-sided pure state black holes, how does a similar construction work? We expect the correlators one needs will depend on the details of the microstates. One natural starting place  might be black holes with an end of world brane like those discussed by Kourkoulou and Maldacena \cite{Kourkoulou:2017zaj}. Finally, while we have focused on charged perturbations, one can equally have bouncing particles in the interior of a rotating black hole \footnote{It is important that the black hole rotates. While test particles with high enough angular momentum can usually bounce outside any kind of black hole, they cannot bounce in the interior of a non-rotating black hole.} by having them carry angular momentum instead. Perturbations with angular momentum have been studied previously in the literature: in rotating BTZ \cite{Reynolds:2016pmi,Malvimat:2021itk} and, very recently, in Kerr-AdS \cite{Malvimat:2022oue}. These results suggest that the scrambling time, as defined through the parameter $\alpha$ in the context of our work, can be made longer by increasing the angular momentum of the perturbation. It would be interesting to understand if this increase in scrambling time is associated to the same features we find in the charged case, namely a delay of the onset of scrambling from the interior perspective due to the presence of a bounce.

\section*{Acknowledgements}

It is a pleasure to thank Felix Haehl, Alexey Milekhin, Diandian Wang and Beni Yoshida for helpful discussions.  G. H., H.L. and L.Q. were supported in part by NSF Grant PHY-2107939. Y.Z. was supported in part by the National Science
Foundation under Grant No. NSF PHY-1748958.

\appendix
\section{Massive charged shell}
\label{app:massive}
In this appendix, we consider the motion of a massive charged shell in a Reissner-Nordstrom AdS background. We then take a massless limit to show that the null trajectory has a turning point. The turning point coincides with $r_b$, where, according to the null matching analysis in section \ref{sec:bounce_inside_bulk}, the shell should bounce in order to have a stress tensor that satisfies the null energy condition.

The equations of motion for charged timelike shells in asymptotically flat black hole backgrounds were previously studied in \cite{1967NCimA..51..744C, 1968CzJPh..18..435K, PhysRevD.8.2363, Hubeny:1998ga} and here we generalize them to the asymptotically AdS case. Similar to the null junction conditions, the conditions for a timelike shell are that the induced metric on the shell is continuous on both sides of the spacetime and that any discontinuity in the extrinsic curvature $K_{ab}$ on the shell implies a surface stress tensor given by
\begin{equation}
    S_{ab}=-\frac{1}{\pi} ([K_{ab}]-[K]h_{ab}),
\end{equation}
where the rectangular brackets denote the difference across the shell. The equations of motion for the shell are obtained by requiring the stress tensor of the shell to take the form of a dust shell $S_{ab}=\sigma u_a u_b$. In the following, we parameterize the shell as $r=R(\tau)$, where $\tau$ is the proper time on the shell. We obtain the equation of motion
\begin{equation}\label{RN_shell}
\frac{m}{R}=\sqrt{f_i(R)+\dot{R}^2}\pm\sqrt{f_o(R)+\dot{R}^2}\,,
\end{equation}
where $m$ is the rest mass of the shell. The sign in \eqref{RN_shell} depends on the sign of the spatial component of the outward unit normal to the shell. Squaring \eqref{RN_shell}, we can rewrite this in the form of a particle in an effective potential
\begin{equation}
    m^2\dot{R}^2+V_{\textnormal{eff}}(R)=0\,,
\end{equation}
where the potential is \begin{equation}
    V_{\textnormal{eff}}(R)=f_i(R)-\left(\frac{R}{2m} (f_i(R)-f_o(R))+\frac{m}{2R}\right)^2\,.
\end{equation}
We can find the turning points of the potential in the limit of small $m$. There is a turning point at
\begin{equation}
    r=l\frac{E}{m}+O(1)\,,
\end{equation}
which is due to the confining nature of asymptotically AdS space and approaches the AdS boundary as we take the massless limit. There are two other turning points at $r=r_b\pm O(m)$, which approach $r_b$ in the massless limit, thus agreeing with the prescription that the null shell is given by two null surfaces joined together at $r_b$.

\section{Scrambling by a charged shell} 
\label{app:scrambling}
In this appendix, we consider the butterfly effect due to a charged shell. As in the case of \cite{Shenker:2013pqa, Leichenauer:2014nxa}, we will see that the scrambling is encapsulated in a shift $\alpha$ in the Kruskal $V$ coordinate along the shell.

Consider a shell that is created on the left boundary at time $t_{wL}$. The initial portion of the trajectory is given by 
\begin{equation}
    U_o=e^{2\pi t_{wL}/\beta_o}\,,\quad U_i=e^{2\pi t_{wL}/\beta_i}
\end{equation}
in terms of Kruskal coordinate $U$ both inside and outside the shell. The $V$ coordinate along this portion of the shell is given by
\begin{equation}
U_o V_o=-e^{4\pi r_o^{*}/\beta_o}\,, \quad U_i V_i=-e^{4\pi r_i^{*}/\beta_i}\,.
\end{equation}
When the shell reaches the turning point $r_b$, the shell will start following a trajectory of constant $V=V_b$. We will first focus on the portion of the shell before the turning point. In the limit that the shell is created at early times $t_0\ll 0$, the shell approaches the horizon. Near the horizon of a non-extremal RN-AdS black hole, we have 
\begin{equation}
r^{*} \approx \frac{\beta}{4\pi}\left(C+\log{\left(\frac{r-r_{+}}{r_{+}-r_{-}}\right)}\right)\,,
\end{equation}
where the constant $C$ depends on the geometry and does not have a compact analytical form in general. Therefore, close to the horizon, we can write
\begin{equation} \label{eq:near_horizon_UV}
U_o V_o \approx -\frac{(r-r_{+}^o)}{r_{+}^o-r_{-}^o}e^{C_o}\,, \quad U_i V_i \approx -\frac{(r-r_{+}^i)}{r_{+}^i-r_{-}^i}e^{C_i}\,.
\end{equation}

The junction conditions require $r$ as a function of $U,V$ to match on both sides. Since we are considering a small perturbation, we can write 
\begin{equation}
C_o=C_i+\delta C\,,\quad r_{+}^o=r_{+}^i+\delta r_+\,, \quad r_{-}^o=r_{-}^i+\delta r_-\,, \quad \beta_o=\beta_i+\delta \beta
\end{equation}
with $\delta C/C_i \sim \delta r_+/r_+^i \sim \delta r_-/r_-^i \sim \delta \beta/\beta_i \ll 1$. Matching $r$ in both sets of coordinates gives the following relationship between the coordinates $V_o$ and $V_i$
\begin{equation}
V_o \approx V_i\left(1+\delta C-\frac{\delta r_{+}-\delta r_-}{r_{+}^i-r_{-}^i}-\frac{2\pi t_{wL}}{\beta_i}\frac{\delta \beta}{\beta_i} \right) + e^{C_i} e^{-2\pi t_{wL}/\beta_i}\frac{\delta r_+}{r_{+}^i-r_-^i}\,,
\end{equation}
while the relationship between $U_o$ and $U_i$ is given by 
\begin{equation}
U_o \approx U_i \left(1-\frac{2\pi t_{wL}}{\beta_i}\frac{\delta \beta}{\beta_i} \right)\,.
\end{equation}
Assuming further that $\beta_i/\delta \beta \gg -t_{wL}/\beta_i \gg 1$, we can write at leading order
\begin{equation}
V_o \approx V_i + \alpha\,,\quad U_o \approx U_i
\end{equation}
with 
\begin{equation}
\alpha=e^{C} e^{-2\pi t_{wL}/\beta} \frac{\delta r_{+}}{r_{+}-r_-}\,,
\end{equation}
where we dropped the $i$ index for the sake of simplicity of notation. 

\section{Post-collision geometry}
\label{app:post_collision}

\begin{figure}[H]
 \begin{center}
        \includegraphics[width=2in]{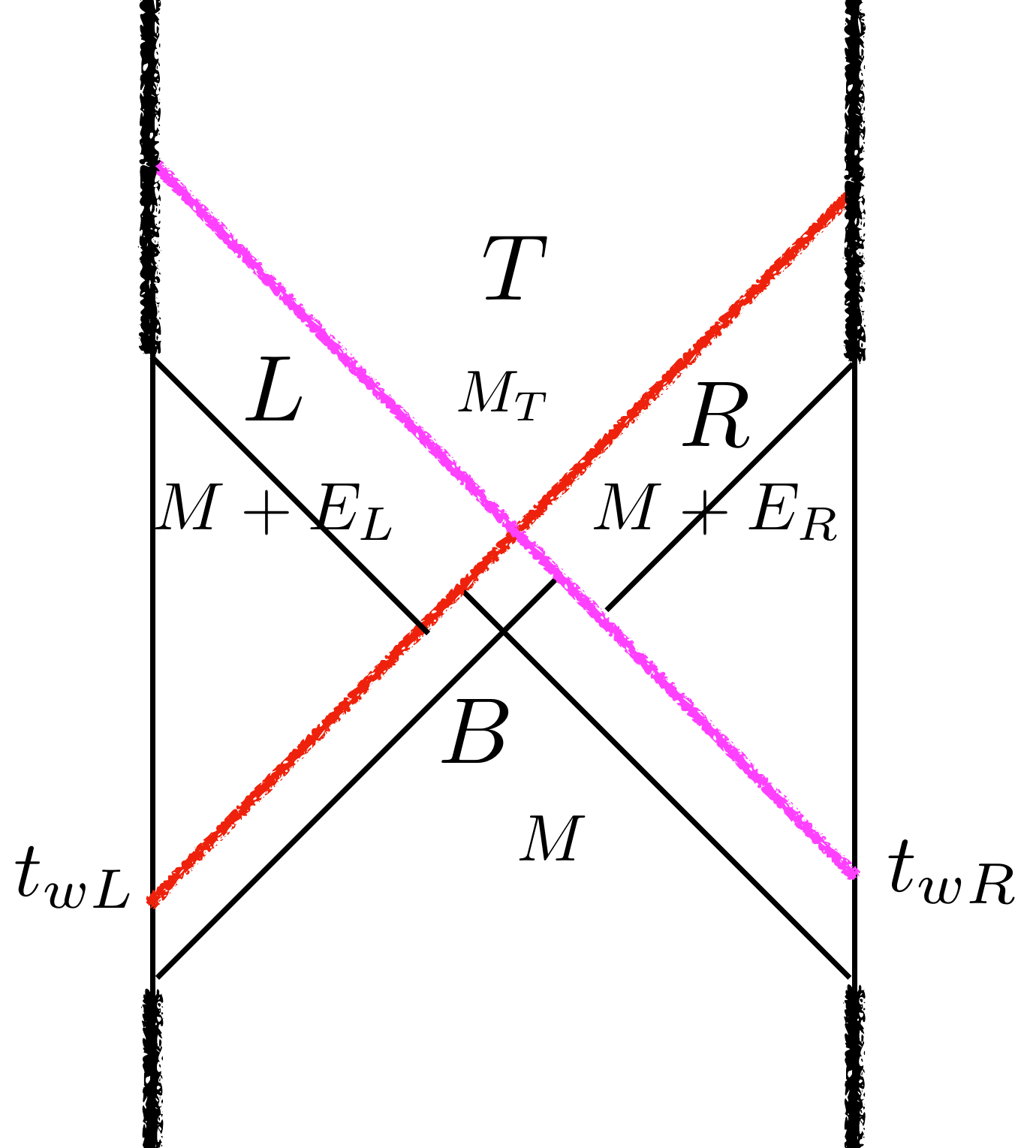}
        \caption{Geometry with two neutral shockwaves colliding}
        \label{Nobouncing_10}
  \end{center}
\end{figure}

In this appendix, we calculate the spacetime volume of the post-collision region for both a collision between two neutral shells and a collision between a charged shell and a neutral shell. The spacetime in the case of two neutral shells is shown in figure \ref{Nobouncing_10}. The shells collide and partition spacetime into 4 regions, labeled by $T,B,L,R$ and each has a metric given by \eqref{RN} with different masses. Inside the black hole, the Kruskal coordinates (in any of the four regions) are related to $r$ by
\begin{equation}\label{inner_tor}
UV=e^{4\pi \Tilde{r}^{*}(r)/ \beta}\,,
\end{equation}
where the inner tortoise coordinate $\Tilde{r}^{*}$ is defined such that $dr^*=dr/f(r)$
and $U,V$ are continuous across the horizon.

The trajectory of the left perturbation is given by $U_{B,l}=e^{2 \pi t_{wL}/\beta_B}$ and that of the right perturbation is given by $V_{B,r}=e^{2 \pi t_{wR}/\beta_B}$ in terms of coordinates in the bottom region. We want to calculate the spacetime volume of the top region, which is given by
\begin{equation}\label{neutral_integral}
    \mathcal{V}=\frac{\beta^2}{2\pi}\int_{U_{T,l}}^\infty dU \int_{V_{T,r}}^\infty dV \frac{r(U,V)^2 \abs{f_T(r(U,V))}}{U V}\,,
\end{equation}
where we have used coordinates in the top region. In order to evaluate \eqref{neutral_integral}, we need to find out where the collision occurs and account for the backreaction from the collision in the top region. Using \eqref{inner_tor} at the collision, one finds that the collision occurs at $r_c$ given by
\begin{equation}
    2\Tilde{r}^*(r_c)=-\Delta t\,,
\end{equation}
where $\Delta t=-t_{wL}-t_{wR}$.
The mass parameters in the $B,L,R$ regions are given by energy conservation. The mass parameter in the top quadrant $M_t$ depends on the collision and is determined by the DTR condition \cite{Dray:1985yt, 10.1143/PTP.73.1401, PhysRevD.41.1796}
\begin{equation}
    f_T(r_c) f_B(r_c)=f_L(r_c) f_R(r_c)\,,
\end{equation}
which gives
\begin{equation}
    M_T(\Delta t)=M+E_L+E_R \frac{f_L(r_c(\Delta t))}{f_B(r_c(\Delta t))}\,.
\end{equation}
Using $M_T$ and \eqref{inner_tor} again at $r_c$ in terms of the $T$ coordinates, one obtains a relation between $U_{T,l}$ and $V_{T,r}$. Since we can use any Kruskal coordinate in the top region, we can fix $U_{T,l}$ arbitrarily and find $V_{T,r}(\Delta t)$
\begin{equation}
    V_{T,r}(\Delta t)=\frac{1}{U_{T,l}}e^{4\pi \Tilde{r}^{*}_T(r_c(\Delta t))/ \beta_T}\,.
\end{equation}
This determines the limits of integration in \eqref{neutral_integral}. The integrand is also determined by $M_T$ and can be obtained from inverting \eqref{inner_tor} in the $T$ coordinates. The integral \eqref{neutral_integral} can then be evaluated numerically.

\begin{figure}[H]
 \begin{center}
        \includegraphics[width=2in]{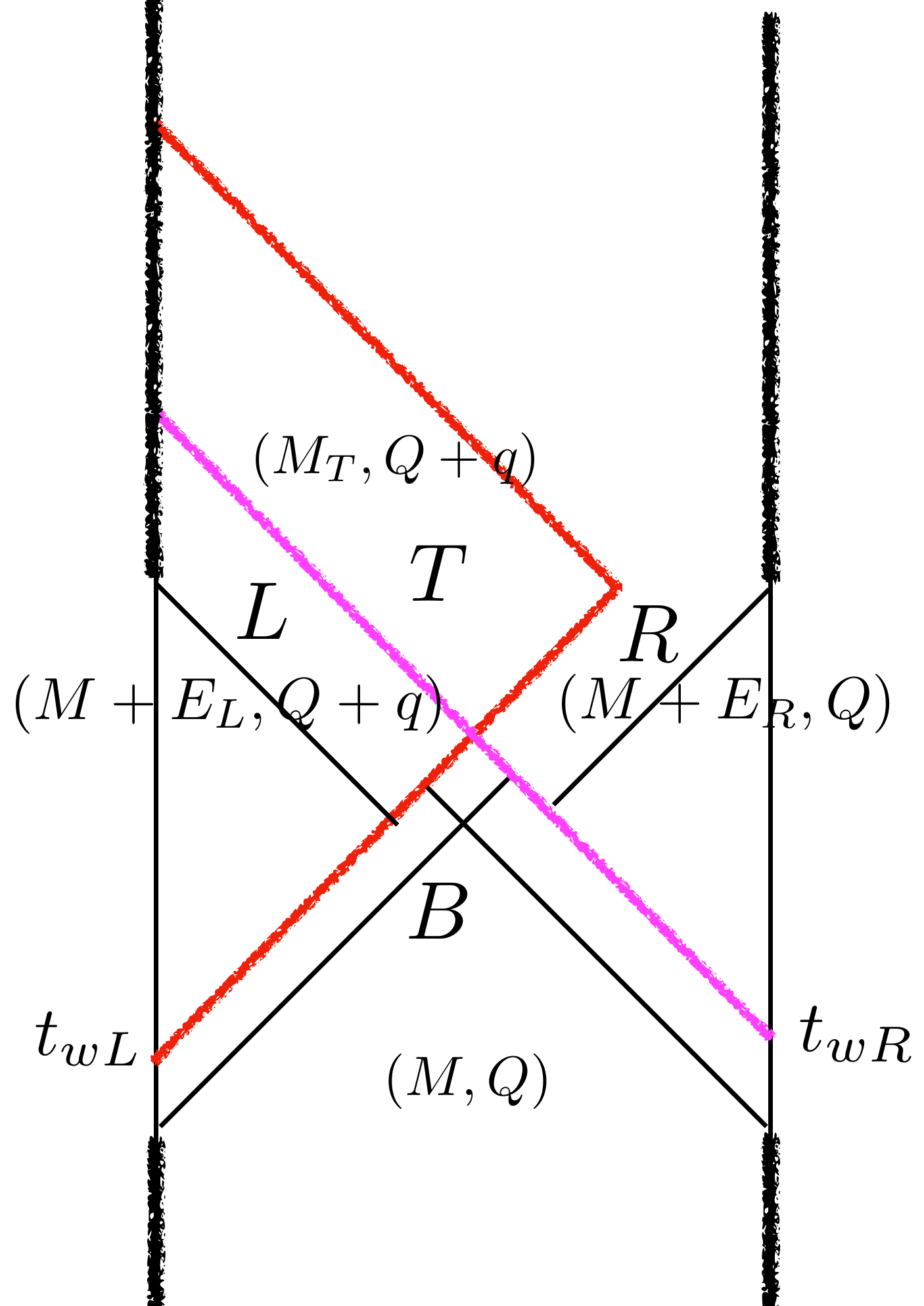}
        \caption{Geometry with a neutral shockwave colliding with a charged shockwave}
        \label{bouncing_20}
  \end{center}
\end{figure}

When a charged shell and a neutral shell collide, the spacetime is as shown in figure \ref{bouncing_20}, where the masses and charges in each region are given by charge and energy conservation and the DTR condition. We will again use Kruskal coordinates. The spacetime volume of the top region is now given by 
\begin{equation}\label{charged_integral}
    V=\frac{\beta^2}{2\pi}\int_{U_{T,l}}^\infty dU \int_{V_{T,r}}^{V_{T,b}} dV \frac{r(U,V)^2 \abs{f_T(r(U,V))}}{U V}\,,
\end{equation}
where we defined $V_{T,b}$ to be the turning coordinate of the shell on the left or $\infty$ if the shell bounces behind the inner horizon ($r_c<R_{T,-}$). One can find the post-collision geometry and the coordinates $U_{T,l}, V_{T,r}$ as before, so it remains to find $V_{T,b}$.

Recall that the turning point $r_b$ is given by
\begin{equation}
    r_b=\frac{q Q_i}{M_o-M_i}\,.
\end{equation}
Here the turning point occurs between the top and right regions. Thus,
\begin{equation}
    r_b(\Delta t)=\frac{q Q}{M_T(\Delta t)-M-E_R}\,.
\end{equation}
Using \eqref{inner_tor} at the turning point, one finds
\begin{equation}
    V_{T,b}(\Delta t)=\frac{1}{U_{T,l}}e^{4\pi \Tilde{r}^{*}_T(r_b(\Delta t))/ \beta_T}
\end{equation}
for $r_t<R_{T,-}$. We then evaluate \eqref{charged_integral} numerically.

\section{Obtaining the geodesic distance in AdS$_2$}
\label{app:ads2_geodesic}
In this appendix, we calculate the geodesic distance between two points, at times $t_L$ and $t_R$ respectively, in each asymptotic boundary of an AdS$_2$ black hole. We compute this geodesic distance taking into account the backreaction of a null particle, sent at time $t_{wL}$ from the left boundary. We assume that this particle can have a turning point somewhere in the interior of the black hole. In order to achieve our goal, we will make use of embedding coordinates. We can describe AdS$_2$ as an embedding 
\begin{equation}
-T_1^2-T_2^2+X^2=-1\,,
\end{equation}
in the geometry
\begin{equation}
ds^2=-dT_1^2-dT_2^2+dX^2\,.
\end{equation}
The appropriate embedding to obtain AdS$_2$ in Kruskal and Schwarzschild-like coordinates in the right asymptotic region is 
\begin{align*}
&T_1=\frac{V+U}{1+UV}=\frac{\beta}{2\pi}\sqrt{(r-r_{+})(r-r_{-})}\sinh{\frac{2\pi t}{\beta}}\,,\\
&T_2=\frac{1-UV}{1+UV}=\frac{2r-r_{-}-r_{+}}{r_{+}-r_{-}}\,,\\
&X=\frac{V-U}{1+UV}=\frac{\beta}{2\pi}\sqrt{(r-r_{+})(r-r_{-})}\cosh{\frac{2\pi t}{\beta}}\,.
\end{align*}
The left asymptotic region can be covered in Schwarzschild-like coordinates by adding $i\beta/2$ to $t$\footnote{Such an analytic continuation yields a coordinate that flows down on the left asymptotic region. We will make it flow up by taking $t\to -t$ after doing the analytic continuation.}. Since our geometry is purely AdS$_2$, we know that the geodesic distance between two spacelike separated points $(T_1,T_2,X)$ and $(T_1',T_2',X')$ is given by 
\begin{equation}
\cosh{d}=T_1 T_1' + T_2 T_2' - XX'\,.
\end{equation}
With this information at hand, we can compute the geodesic distance between two points in each asymptotic boundary. In order to do that, since we are taking into account the backreaction of a null particle that bounces somewhere along its trajectory, we first need to consider that the geodesic might pass through the particle either before or after it bounces. Let us first consider connecting both boundary points by a geodesic that meets the shell before it bounces, i.e. it meets the shell at a point $(U_w,V_m \leq V_b)$. Using the embedding coordinates, we can compute the total distance $d=d_1+d_2$ between the two boundaries along curves which are geodesics until meeting the shell. Here, $d_1$ is the geodesic distance from the left asymptotic boundary $(t_L,r_c)$ to a point $V$ along the trajectory of the particle and $d_2$ is the geodesic distance from there to the right asymptotic boundary $(t_R,r_c)$. We find in the large $r_c$ limit 
\begin{equation}\label{geodesic_distance_V}
\begin{split}
e^{d}&=\frac{\beta \tilde{\beta}^2r_c^2 e^{-\frac{2\pi t_L}{\tilde{\beta}}} e^{\frac{2\pi t_R}{\beta}}}{4\pi^3}\frac{\left(e^{-\frac{2\pi t_R}{\beta}}+U_w\right)\left(e^{\frac{2\pi t_L}{\tilde{\beta}}}-\tilde{U}_w\right)\left(1-e^{-\frac{2\pi t_R}{\beta}}V\right)}{\tilde{U}_w(1+U_wV)^2}\cdot\\
&\cdot \left(e^{\frac{2\pi t_L}{\tilde{\beta}}}(\tilde{r}_{+}-r_{+}+(\tilde{r}_{+}-r_{-})U_wV)+\tilde{U}_w(r_{+}+r_{-} U_wV-\tilde{r}_{-}(1+U_wV) \right)\,.
\end{split}
\end{equation}
Not all gluings of such pairs of geodesics are themselves a geodesic of the perturbed geometry. To find the actual geodesic distance between the two boundaries, we must extremize $d$ with respect to $V$. In other words, we need to find the point $V=V_m$ at which the gluing actually yields the geodesic of the perturbed geometry connecting both sides. We find a maximum at
\begin{equation}\label{V_meet}
V_m=\frac{1}{U_w}\frac{(r_{+}-\tilde{r}_{-})-(r_{-}+\tilde{r}_{-}-2r_{+})e^{\Delta t_R}+\delta r_{+} e^{\Delta t_L}+(r_{-}-r_{+}+\delta r_{+})e^{\Delta t_L}e^{\Delta t_R}}{(e^{\Delta t_L}-1)(2+e^{\Delta t_R})r_{-}+(\tilde{r}_{-}+r_{+})+\tilde{r}_{-}e^{\Delta t_R}-e^{\Delta t_L}(r_{+}+\tilde{r}_{+}+\tilde{r}_{+}e^{\Delta t_R})}
\end{equation}
where we defined 
\begin{equation}
\delta r_{+}=\tilde{r}_{+}-r_{+}\,,\,\,\, \delta r_{-}=\tilde{r}_{-}-r_{-}\,
\end{equation}
and
\begin{equation}
\Delta t_{L}=\frac{2\pi}{\tilde{\beta}}(t_{L}-t_{wL})\,,\,\,\, \Delta t_{R}=\frac{2\pi}{\beta}(-t_{wL}-t_R)\,.
\end{equation}
Evaluating eq. \eqref{geodesic_distance_V} at $V=V_m$ yields
\begin{equation}\label{geodesic_distance_app}
\begin{split}
e^{-L}&=\frac{64\pi^4 e^{\Delta t_L}e^{\Delta t_R}}{\beta^2 \tilde{\beta}^2 r_c^2 \left[\delta r_{-}+(\tilde{r}_{-}-r_{+})e^{\Delta t_R}+(r_{-}-\tilde{r}_{+})e^{\Delta t_L}-\delta r_{+}e^{\Delta t_L}e^{\Delta t_R}\right]^2}\,,
\end{split}
\end{equation}
where $L$ is the geodesic distance. A priori, this computation is only valid for boundary times such that $V_m \leq V_b$. Afterwards, the geodesic will start meeting the particle at a point $(U_m \geq U_w,V_b)$. We can follow an analogous procedure to find the geodesic distance $L'$ in this case. The expression turns out to be the same as \eqref{geodesic_distance_app} if the DTR condition is satisfied, i.e. if we impose eqs. \eqref{r_b_eq} and \eqref{V_b}. In this case we also find that $V_m=V_b \iff U_m=U_w$, i.e. as we move the boundary times, the geodesic smoothly interpolates between the constant $U_w$ and constant $V_b$ parts of the particle trajectory. In summary, when the DTR condition is satisfied, the geodesic distance is always given by \eqref{geodesic_distance_app} and as a consequence it is completely smooth across the turning point as we change the boundary times.

%%%%%%%%%%%%%%%%%%%%%%%%%%%%%%%%%%%%%%%%%%%%%%
%%%%%%%%%%%%%%%%%%%%%%%%%%%%%%%%%%%%%%%%%%%%%%
%~~~~~~~~~~~~~~~~~~~~~~~~~~~~~~~~~~~~~~~~~~~~~~
\bibliographystyle{utphys}
\bibliography{reference}
%~~~~~~~~~~~~~~~~~~~~~~~~~~~~~~~~~~~~~~~~~~~~~~

\end{document}